\documentclass[a4paper,onecolumn,unpublished,amsfonts,amsmath,amssymb,noarxiv]{quantumarticle}
\pdfoutput=1
\usepackage[utf8]{inputenc}
\usepackage[english]{babel}
\usepackage[T1]{fontenc}
\usepackage[unicode=true,
bookmarks=true,bookmarksopen=false,
breaklinks=false,pdfborder={0 0 0},colorlinks=true]
{hyperref}
\usepackage{xcolor}
\definecolor{cblue}{rgb}{0.16, 0.32, 0.75}
\definecolor{cred}{rgb}{0.7, 0.11, 0.11}
\hypersetup{%
	,linkcolor=cred
	,citecolor=cblue
	,urlcolor=cblue
}
\usepackage[numbers,sort&compress]{natbib}
\usepackage{dsfont,enumitem,color,bm,ulem,comment,cancel}
\usepackage[allowbreakbefore,nospacearound,shortcuts]{extdash}

\newcommand{\fbar}[1]{
    \,\overline{\!#1}
}
\newcommand{\fbarsb}[2]{
    {\,\overline{\!#1}}{\vphantom{#1}}_{\!#2}
}
\newcommand{\sqb}[1]{[\,#1^+\!\!,#1^-]}
\newcommand{\sqbc}[2]{[\,#1^+\!\!,#1^-|\,#2^+\!\!,#2^-]}

\newcommand{\bimeasure}[2][]{
    \mathcal{Q}^{#1}[\,#2^+\!\!,#2^-][\mathcal{D}#2^+][\mathcal{D}#2^-]
}
\newcommand{\tup}[2][n]{
    \bm{#2}_{#1}
}

\renewcommand{\part}[1]{
   ~\refstepcounter{part}
    
    \noindent{\Large\sffamily \textbf{Part \Roman{part}}}
    
    \vspace{1em}
    \noindent{\huge\sffamily #1}
    
    \vspace{0.5em}
}

\numberwithin{equation}{section}
\allowdisplaybreaks		

\usepackage{amsthm}
\theoremstyle{definition}

\begin{document}

\title{Phenomenological quantum mechanics: \newline II. Deducing the formalism from experimental observations}
\author{Piotr Sza{\'n}kowski}
\orcid{0000-0003-4306-8702}
\address{Institute of Physics, Polish Academy of Sciences, al.~Lotnik{\'o}w 32/46, PL 02-668 Warsaw, Poland}
\email{piotr.szankowski@ifpan.edu.pl}

\author{Davide Lonigro}
\orcid{0000-0002-0792-8122}
\address{Department Physik, Friedrich-Alexander-Universität Erlangen-Nürnberg, Staudtstraße 7, 91058 Erlangen, Germany}
% \address{Dipartimento di Matematica, Università degli Studi di Bari Aldo Moro, via E. Orabona 4, 70125 Bari, Italy}
% \address{Istituto Nazionale di Fisica Nucleare, Sezione di Bari, via G. Amendola 173, 70126 Bari, Italy}
%\email{email}

\author{Fattah Sakuldee}
\orcid{0000-0001-8756-7904}
\address{Wilczek Quantum Center, School of Physics and Astronomy, Shanghai Jiao Tong University, 800 Dongchuan Road, Minhang, 200240 Shanghai, China}
\address{The International Centre for Theory of Quantum Technologies, University of Gda\'nsk, Jana Ba\.zy\'nskiego 1A, 80-309 Gda\'nsk, Poland}
%\email{fattah.sakuldee@sjtug.edu.cnpl}

\author{{\L}ukasz Cywi{\'n}ski}
\orcid{0000-0002-0162-7943}
\address{Institute of Physics, Polish Academy of Sciences, al.~Lotnik{\'o}w 32/46, PL 02-668 Warsaw, Poland}
%\email{lcyw@ifpan.edu.pl}

\author{Dariusz Chru\'{s}ci\'{n}ski}
\orcid{0000-0002-6582-6730}
\address{Institute of Physics, Faculty of Physics, Astronomy and Informatics, Nicolaus Copernicus University,
Grudziadzka 5/7, 87-100 Toru\'n, Poland}
%\email{darch@fizyka.umk.pl}

\begin{abstract}
We propose an exercise in which one attempts to deduce the formalism of quantum mechanics solely from phenomenological observations. The only assumed inputs are the multi-time probability distributions estimated from the results of sequential measurements of quantum observables; no presuppositions about the underlying mathematical structures are permitted.

In the concluding Part II of the paper, we carry out the deduction of the formalism from the phenomenological inputs described in Part~I. We show that the resulting formalism exhibits an affinity with Hilbert spaces, and we derive an explicit representation in terms of those mathematical structures. Analogues of the obtained elementary building blocks---such as projection operators---are readily identifiable within the standard formalism. However, once these building blocks are assembled according to the blueprint of the deduced bi-trajectory formalism, it becomes evident that the new and the standard formalisms differ substantially at the conceptual level. 

These differences do not negate the fact that both formalisms are in perfect agreement with respect to empirically testable predictions. Rather, the emergence of a novel, non-standard formulation should be seen as a relatively rare opportunity to reassess, from a fresh perspective, some of the long-standing foundational issues in the theory. The hope is that the new approach may prove more successful in addressing problems that have resisted resolution within the established theoretical framework.
\end{abstract}

%\date{\today}

\maketitle

\tableofcontents

\section{Introduction}\label{sec:intro}

In the first part of this paper~\cite{part1}, the analysis of the phenomenology of sequential quantum measurements have lead us to the conclusion that the emerging formalism for the quantum theory should be based on the \textit{bi-trajectory} formulation. Having established the overarching formal framework, we now proceed to deduce its representation in terms of Hilbert space structures. As expected from the formalism of quantum mechanics, the connection with Hilbert spaces is not accidental; we shall demonstrate that it follows naturally from the fundamental properties of the bi-probability distributions we have chosen as a basis for the formalism. However, the precise form of the representation cannot be deduced from the abstract mathematical properties of bi-probabilities alone. For the deduction to be successful, it must be informed by the phenomenological observations collected in Part I.

Finding the Hilbert space representation is a crucial step: it will enable us to disassemble the elements of the bi-trajectory formalism into their ``constituent parts''. These elementary building blocks of the theory can then be analyzed, understood, and, most importantly, purposefully reassembled into new forms that have not been seen before, or were not necessarily accessible through phenomenological observations. This kind of ability is essential for ensuring the theory's predictive and explanatory power. A formalism limited to merely describing experimental results it has been deduced from, without the capacity to model and describe \textit{potential} systems and experimental scenarios that have not yet been observed, would be incomplete.

\section{Outline}

We begin in Section~\ref{sec:part1} with a brief summary of Part I: Section~\ref{sec:pheno_prob_dist} recalls the definitions, the notation conventions, and the phenomenological observations regarding the multi-time probability distributions describing sequential quantum measurements. Next, Section~\ref{sec:bi-prob_dist} presents the previously deduced basis elements of the formalism: the bi-probability distributions.

In Section~\ref{sec:Gudder_inner_prod}, we establish the critical link between bi-probabilities and Hilbert spaces via Gudder's theorem on inner product representations~\cite{Gudder_MathSlov12}. In Section~\ref{sec:obs--projector_link}, we exploit this connection along with the phenomenology of single-measurement experiments to infer the measurement--projector link, a foundational concept in the standard formalism. In Section~\ref{sec:initialization_event} we deduce the Hilbert space representation of the initialization event by combining the projector link and the Markovianity property of multi-time probability distributions. Finally, in Section~\ref{sec:H-rep_for_bi-probs}, we derive how bi-probabilities are represented in Hilbert space structures, such as projector operators. In the process, we also reveal the limitations of the measurement--projector link.

In Section~\ref{sec:dynamical_laws}, we further refine the Hilbert space representation, drawing on the phenomenology of uncertainty relations and the Zeno effect, leading to the appearance of time-dependent unitary operators---and their generators---which are identified as the formal representation of the quantum dynamical laws.

In Section~\ref{sec:composite_sys}, we infer the form of the Hilbert space corresponding to a system composed of independent subsystems. We then deduce that independent systems become dependent when the dynamical law includes \textit{interaction terms}. This establishes the correspondence between Hermitian operators and observables measurable by devices deployed by an observer.

Finally, Section~\ref{sec:ture_master} concludes the second part of the paper by identifying the \textit{master object} of the bi-trajectory formalism---the fundamental element from which all other elements, including phenomenologically accessible multi-time probabilities, are derived.
Final considerations and outlooks are gathered in Section~\ref{sec:conclusions}.

\section{Summary of Part I}\label{sec:part1}

\subsection{Phenomenological multi-time probability distributions}\label{sec:pheno_prob_dist}

The experiments considered in Part I consisted of a sequential probing the system's observables 
\begin{align}
\tup{F} = (F_n,\ldots,F_1)
\end{align}
by deploying at premeditated successive times 
\begin{align}
\tup{t} = (t_n,\ldots,t_1);\quad t_n>\cdots>t_1
\end{align}
specific measuring devices---$F_j$-devices, as we called them---to obtain a sequence of outcomes
\begin{align}
    \tup{f} = (f_n,\ldots,f_1)\in\Omega(\tup{F})=\Omega(F_n)\times\cdots\times\Omega(F_1),
\end{align}
where $\Omega(F_j)$ is the finite set of possible results produced by the corresponding device. These outcomes, together with the information about the initial condition, were used to estimate the multi-time probability distributions
\begin{align}
    P^{F_n,\ldots,F_1|p}_{t_n,\ldots,t_1|t_0}(f_n,\ldots,f_1) = P^{\tup{F}|p}_{\tup{t}|t_0}(\tup{f}),
\end{align}
which collectively form the full phenomenological description of each experiment.

The first conclusion drawn from our analysis was that sequential quantum measurements do exhibit the standard notion of causality. We have found that the past measurement outcomes may affect the future outcomes, but not the other way around; formally, this property is encapsulated by the result for the conditional probability of measuring an outcome given the previously measured sequence:
\begin{align}
\label{eq:conditional_prob}
    P^{F_n|\tup[n-1]{F}}_{t_n|\tup[n-1]{t}}(f_n|\tup[n-1]{f})
    &= \frac{P^{\tup{F}|p}_{\tup{t}|t_0}(\tup{f})}{P^{\tup[n-1]{F}|p}_{\tup[n-1]{t}|t_0}(\tup[n-1]{f})},
\end{align}
which follows from the observed relation between probability distributions for sequences of different lengths
\begin{align}
    \text{\textbf{Phenomenological observation} (Causality)}:\quad
    \sum_{f_n\in\Omega(F_n)}P^{\tup{F}|p}_{\tup{t}|t_0}(\tup{f})
    = P^{\tup[n-1]{F}|p}_{\tup[n-1]{t}|t_0}(\tup[n-1]{f}).
\end{align}
By no means this is a trivial statement as, in general, the phenomenological probabilities were found to violate the classical consistency condition:
\begin{align}\label{obs:inconsistency}
    \begin{array}{r}
    \text{\textbf{Phenomenological}}\\
    \text{\textbf{observation}}\\
    \text{(Inconsistency)}
    \end{array}:\ 
    \sum_{f_j\in\Omega(F_j)}P^{\tup{F}|p}_{\tup{t}|t_0}(\tup{f}) \neq
    P^{
        F_n,\ldots,\cancel{F_j},\ldots,F_1|p
    }_{
        t_n,\ldots,\cancel{t_j},\ldots,t_1|t_0
    }(f_n,\ldots,\cancel{f_j},\ldots,f_1)\quad\text{for $j<n$}.
\end{align}
Hence, the fact that the marginalization over the latest entry in the sequence specifically always leads to the probability for a shorter sequence, is significant.

The degree of ``graining'' of the measuring device is one of the crucial concepts introduced in Part~I. An $\fbar{F}$-device is considered a coarse-grained version of an $F$-device, when both devices measure the same observable $F$ but the coarse-grained variant is incapable of distinguishing between some of the results available to the $F$-device; in formal terms, the relationship between the devices is quantified with the resolution of $\fbar{F}$-device with respect to $F$-device:
\begin{align}\label{eq:resolution}
\operatorname{Res}(\fbar{F}\mid F) = \Bigl\{\omega(\fbar{f})\subseteq\Omega(F)\ \Bigm|\   \fbar f\in\Omega(\fbar{F}),\ \forall_{\fbar{f}\neq\fbar{f}'}\omega(\fbar{f})\cap\omega(\fbar{f}')=\emptyset,\ \bigcup_{\fbar f\in\Omega(\fbar{F})}\omega(\fbar f) = \Omega(F)\Bigr\};
\end{align}
additionally, we have introduced the notation convention that allows to designate the resolution of a device (at least partially) directly on the level of probability distributions:
\begin{align}
    P^{\fbarsb{\bm{F}}{n}|p}_{\tup{t}|t_0}(\fbarsb{\bm f}{n}) 
    =P^{\fbarsb{\bm F}{n}|p}_{\tup{t}|t_0}\Big(
        \bigvee_{\omega(\fbarsb{\bm f}{n})}\tup{f}
    \Big)
    = P^{\fbarsb{\bm F}{n}|p}_{\tup{t}|t_0}\Big(
        \bigvee_{\omega(\fbarsb{f}{n})}f_n,\ldots,\bigvee_{\omega(\fbarsb{f}{1})}f_1    
    \Big).
\end{align}
In this sense, a measuring device counts as \textit{perfectly fine-grained} when it is \textit{not} a coarse-grained variant of any other measuring device. By convention, we denote the observables measured by such devices with $K,L,\ldots$, while leaving $F,G,\ldots$ to indicate generic observables of unspecified degree of graining.

The use of perfectly fine-grained devices was indispensable for resolving the issue of the \textit{initialization event}. Based on our analysis, we have determined that the considered experiments can always be initialized by deploying the devices of the same kind as those used to probe system's observables, provided that the initializing device is perfectly fine-grained:
\begin{align}\label{eq:phenomenology_of_initialization}
    P^{\tup{F}|p}_{\tup{t}|t_0}(\tup{f}) = \sum_{K}\sum_{k\in\Omega(K)}P_{\tup{t}|t_0}^{\tup{F}|K}(\tup{f}|k)p^K(k)\quad\text{such that $\sum_{K,k} p^K(k) = 1,\,p^K(k)\geq 0$},
\end{align}
where $P^{\tup{F}|K}_{\tup{t}|t_0}(\tup{f}|k)$ indicates the probability of measuring the sequence $\tup{f}$ at times $\tup{t}$ with $\tup{F}$-devices, given the outcome $k\in\Omega(K)$ of the measurement with a perfectly fine-grained $K$-device deployed at the initial time $t_0 < t_1$.

The relation~\eqref{eq:phenomenology_of_initialization} follows from the observed factorization of multi-time probabilities describing the sequential deployment of perfectly fine-grained $K_j$-devices:
\begin{align}\label{obs:markov}
    \begin{array}{r}
        \text{\textbf{Phenomenological}}\\
        \text{\textbf{observation}}\\
        \text{(Markovianity)}
    \end{array}:\quad 
    P^{\bm K_n|p}_{\tup{t}|t_0}(\bm{k}_n) &= P^{K_n|K_{n-1}}_{t_n|t_{n-1}}(k_n|k_{n-1})\cdots P^{K_2|K_1}_{t_2|t_1}(k_2|k_1)P^{K_1|p}_{t_1|t_0}(k_1),
    \end{align}
where the factors $P^{K_j|K_{j-1}}_{t_j|t_{j-1}}(k_j\mid k_{j-1})$ correspond to conditional probabilities in the sense of definition~\eqref{eq:conditional_prob}. The implication is that the outcomes measured after the deployment of a perfectly fine-grained device depend on its result but, crucially, do not depend on the results obtained before. Then, the outcomes measured prior to the deployment of the device can be disregarded because measurements taken afterwards no longer depend on those previous measurement results. In other words, the deployment of a perfectly fine-grained device cuts off the past events from influencing the future---an adequate definition of initialization event.

Measurements with perfectly fine-grained devices were also used to demonstrate the second---after the inconsistency relation~\eqref{obs:inconsistency}---point of divergence between the phenomenology of classical and quantum systems. As it was explained in Part I, a system is classical when it can be described with a uni-trajectory theory, where any system's observable $F^\mathrm{cl}$ is represented by a trajectory $t\mapsto f(t)$ with a probability measure $\mathcal{P}^{F^\mathrm{cl}}[\,f\,][\mathcal{D}f]$. The existence of the master measure $\mathcal{P}^{F^\mathrm{cl}}$ requires that multi-time probabilities describing sequential measurements satisfy consistency conditions, and thus, the inconsistency relation~\eqref{obs:inconsistency} indicates that the systems investigated in Part I are incompatible with any classical uni-trajectory theory. Another key implication of the uni-trajectory picture is the \textit{equivalence} of classical observables: trajectories representing any classical observable are functions of the one elementary observable, and consequently, outcomes of perfectly fine-grained measurements contain the same information as the measurements of the elementary observable. The results of an experiment where perfectly fine-grained measuring devices were deployed in a rapid succession demonstrated that this is not the case for systems investigated in Part I:
\begin{align}\label{obs:uncertainty}
    \begin{array}{r}
    \text{\textbf{Phenomenological}}\\
    \text{\textbf{observation}}\\
    \text{(Uncertainty relations)}
    \end{array}:\ 
    \lim_{\Delta t\to 0^+} P^{K|L}_{t+\Delta t|t}(k|\ell) 
        = \lim_{\Delta t\to 0^+} P^{L|K}_{t+\Delta t|t}(\ell|k) \equiv C^{K|L}_{k,\ell},\quad\text{for any $t>0$,}
\end{align}
where the classical-like outcome $C^{K|L}_{k,\ell} = \delta_{k,h(\ell)}$ (with a one-to-one function $h:\Omega(L)\to\Omega(K)$) is obtained when $K$ and $L$ are measured by an equivalent perfectly fine-grained device, but, in general, it is possible to find two devices resulting in $C^{K|L}_{k,\ell}\neq \delta_{k,h(\ell)}$. Therefore, quantum systems can support \textit{inequivalent} observables---something inconceivable in classical theories.

Nevertheless, even experiments with equivalent quantum observables can still reveal the non-classical nature of the investigated systems. In particular, the experiment where a single $K$-device is deployed with increasing frequency over a fixed period $t$, leads to the following result: 
\begin{align}\label{obs:zeno}
    \begin{array}{r}
    \text{\textbf{Phenomenological}}\\
    \text{\textbf{observation}}\\
    \text{(Quantum Zeno effect)}
    \end{array}:\ 
    \lim_{n\to\infty} P^{K|K}_{\bm{s}_n|0}(k_0,\ldots,k_0|k_0) 
        = \lim_{n\to\infty}\prod_{j=0}^{n-1} P^{K|K}_{s_j + \frac{t}{n}|s_j}(k_0|k_0) = 1
    \,;\quad s_j = \frac{jt}{n}.
\end{align}
We have argued in Part I that, aside being fundamentally incompatible with the classical uni-trajectory picture, the observed Zeno effect gives us important insight into the short-time behavior of the quantum dynamical laws. 

Next, we have proceeded to analyze the phenomenology of experiments involving coarse-grained measuring devices. In systems described by classical theories, due to observable equivalence, coarse-grained measurement device can always be represented by a fine-grained device that post-processes the displayed result; therefore, the readout of a classical coarse-grained $\fbarsb{F}{j}^\mathrm{cl}$-device is equivalent to an \textit{alternative} of readouts with the corresponding fine-grained classical $F_j^\mathrm{cl}$-device,
\begin{align}
    P^{\ldots\fbarsb{F}{j}^\mathrm{cl}\ldots}_{\ldots t_j \ldots}\Big(
        \ldots\bigvee_{\omega(\fbarsb{f}{j})}f_j\ldots
    \Big) = \sum_{f_j\in\omega(\fbarsb{f}{j})}P^{\ldots F_j^\mathrm{cl}\ldots}_{
        \ldots t_j \ldots
    }(\ldots f_j \ldots).
\end{align}
This turns out not to be the case for non-classical systems investigated in Part I: it is possible to construct a quantum coarse-grained $\fbarsb{F}{j}$-device probing observable $F_j$, such that when it is deployed mid sequence in place of fine-grained $F_j$-device, the obtained outcomes subvert the classical picture,
    \begin{align}%\label{eq:resolution_crit}
        \begin{array}{r}
        \text{\textbf{Phenomenological}}\\
        \text{\textbf{observation}}\\
        \text{(Quantum coarse-grained}\\
        \text{measurement)}
        \end{array}:\ 
        P_{t_n,\ldots,t_j,\ldots,t_1|t_0}^{F_n,\ldots,\fbarsb{F}{j},\ldots,F_1|p}\Big(
            f_n,\ldots,\bigvee_{\omega(\fbarsb{f}{j})}f_j,\ldots,f_1\Big) 
        \neq \sum_{f_j\in\omega(\fbarsb{f}{j})}P_{\tup{t}|t_0}^{\tup{F}|p}(\tup{f}),
    \end{align}
    However, when the same kind of device is deployed at the end of the sequence, the classical picture always persists:
    \begin{align}\label{obs:coarse-grained_measurement}
        \begin{array}{r}
        \text{\textbf{Phenomenological}}\\
        \text{\textbf{observation}}\\
        \text{(The terminal coarse-grained}\\
        \text{measurement)}
        \end{array}:\ 
        P^{\fbarsb{F}{n},F_n,\ldots,F_1|p}_{t_n,t_{n-1},\ldots,t_1|t_0}\Big(\bigvee_{\omega(\fbarsb{f}{n})}f_n,f_{n-1},\ldots,f_1\Big)
            = \sum_{f_n\in\omega(\fbarsb{f}{n})}P^{\tup{F}|p}_{\tup{t}|t_0}(\tup{f}).
    \end{align}

Experiments involving quantum coarse-grained measuring devices can be understood as an implementation of a sequential generalization of $k$-slit experiment probing the quantum interference effects.

\subsection{Bi-probability distributions}\label{sec:bi-prob_dist}
A number of observations made in Part I have been accounted for in the initial stage of the deduction. There, we used some of the collected phenomenological inputs---mainly those referring to the phenomenon of quantum interference---to introduce the bi-probability distributions $Q^{\tup{F}|p}_{\tup{t}|t_0}$ to constitute the basis for the emerging formalism.

Formally, the bi-probabilities are complex-valued distributions on the space of sequence pairs,
\begin{align}
    Q^{\tup{F}|p}_{\tup{t}|t_0}:\ \Omega(\tup{F})\times\Omega(\tup{F})\to \mathbb{C},
\end{align}
and are defined by the following list of properties:
\begin{enumerate}[label=\textnormal{(Q\arabic*)}]
    \item\label{prop:bi-prob:norm} \textit{Normalization}
        \begin{align*}
            \sum_{\tup{f}^\pm\in\Omega(\tup{F})}Q^{\tup{F}|p}_{\tup{t}|t_0}(\tup{f}^+,\tup{f}^-) = 1    
        \end{align*}
    \item\label{prop:bi-prob:causality} \textit{Causality}:
        \begin{align*}
            Q^{F_n,\ldots,F_1|p}_{t_n,\ldots,t_1|t_0}(f_n^+,\ldots, f_1^+\,;\,f_n^-,\ldots,f_1^-) \propto \delta_{f_n^+,f_n^-};
        \end{align*}
    \item\label{prop:bi-prob:factorization}\textit{Factorization rule for independent systems}:
        \begin{align*}
            Q^{F_AF_B}_{\tup{t}}(\bm{a}^+_n\!\wedge\bm{b}_n^+ , \bm{a}_n^-\!\wedge\bm{b}_n^-)
                = Q^{F_A}_{\tup{t}}(\bm{a}_n^+,\bm{a}_n^-)Q^{F_B}_{\tup{t}}(\bm{b}_n^+,\bm{b}_n^-);
        \end{align*}
        where $F_AF_B$ indicates the observable measured by the simultaneous deployment of the $F_A$- and $F_B$-devices that measure observables belonging to the independent systems $A$ and $B$, respectively.
    \item\label{prop:bi-prob:pos} \textit{Positive semi-definiteness}:
        \begin{align*}
            \sum_{\tup{f}^\pm\in\Omega(\tup{F})}
                Z(\tup{f}^+)
                Q^{\tup{F}|p}_{\tup{t}|t_0}(\tup{f}^+,\tup{f}^-)
                Z(\tup{f}^-)^* \geq 0
                \quad\text{for any function $\tup{f}\mapsto Z(\tup{f})\in\mathbb{C}$};
        \end{align*}
    \item\label{prop:bi-prob:bi-consistency} \textit{Bi-consistency}:
        \begin{align*}
            \sum_{f_j^\pm\in\Omega(F_j)}Q^{\tup{F}|p}_{\tup{t}|t_0}(\tup{f}^+,\tup{f}^-) = Q^{F_n,\ldots,\cancel{F_j},\ldots,F_1|p}_{t_n,\ldots,\cancel{t_j},\ldots,t_1|t_0}(f_n^+,\ldots,\cancel{f_j^+},\ldots,f_1^+\,;\,f_n^-\ldots,\cancel{f_j^-},\ldots,f_1^-);
        \end{align*}
    \item\label{prop:bi-prob:master_measure} \textit{The bi-trajectory picture}: The bi-probabilities $Q^{F|p}_{\tup{t}|0}$ associated with a single observable $F$ are discrete-time restrictions of a complex-valued measure $\bimeasure[F|p]{f}$ on the space of bi-trajectories, $t\mapsto (f^+(t),f^-(t))\in\Omega(F)\times\Omega(F)$:
    \begin{align*}
        Q^{F|p}_{\tup{t}|0}(\tup{f}^+,\tup{f}^-) 
        &= \iint \Big(\prod_{j=1}^n\delta_{f^+(t_j),f^+_j}\delta_{f^-(t_j),f_j^-}\Big)\bimeasure[F|p]{f};
    \end{align*}
    \item\label{prop:bi-prob:measurement_link} \textit{Measurement link}: the phenomenological probability distributions are equal to the diagonal part of the corresponding bi-probability:
        \begin{align*}
            Q^{\tup{F}|p}_{\tup{t}|t_0}(\tup{f},\tup{f}) = P^{\tup{F}|p}_{\tup{t}|t_0}(\tup{f});
        \end{align*}
    \item\label{prop:bi-prob:additivity} \textit{Quantum interference}: bi-probability distributions associated with coarse-grained observables exhibit additivity with respect to coarse-graining $\vee$ treated as an operation on sequences:
        \begin{align*}
        Q^{\fbarsb{\bm{F}}{n}|p}_{\tup{t}|t_0}\Big(\bigvee_{\omega(\fbarsb{\bm{f}}{n})}\tup{f}^+,\bigvee_{\omega(\fbarsb{\bm{f}}{n})}\tup{f}^-\Big)
            = \sum_{\tup{f}^\pm\in\omega(\fbarsb{\bm{f}}{n})} Q_{\tup{t}|t_0}^{\tup{F}|p}(\tup{f}^+,\tup{f}^-).
        \end{align*}
\end{enumerate}

\section{The Hilbert space representation}\label{sec:H-space_rep}

\subsection{Bi-probability distributions as inner products}\label{sec:Gudder_inner_prod}

Consider an abstract vector space over complex numbers $\mathbb{C}^d$, with the dimension $d = |\Omega(\tup{F})|$ being equal to the number of unique sequences $\tup{f}\in\Omega(\tup{F}) = \Omega(F_n)\times\cdots\Omega(F_1)$. Then, choose an arbitrary orthonormal basis, and enumerate its elements with sequences:
\begin{align}
    \big\{ |\psi(\tup{f}) \rangle \in \mathbb{C}^d \ \big|\  \tup{f}\in\Omega(\tup{f}),\,\langle\psi(\tup{f}^+)|\psi(\tup{f}^-)\rangle = \delta_{\tup{f}^+,\tup{f}^-}
    \big\},
\end{align}
where $\langle\,\cdot\,|\,\cdot\,\rangle: \mathbb{C}^d\times\mathbb{C}^d\to\mathbb{C}$ is the standard inner product.

Since the bi-probability $Q^{\tup{F}|p}_{\tup{t}|t_0}$ is a positive semi\-/definite bi\-/sequence distribution (see property~\ref{prop:bi-prob:pos}), any bi\-/linear form defined as
\begin{align}
    \langle\!\langle \Psi | \Phi \rangle\!\rangle_{\tup{t}|t_0}^{\tup{F}|p} :=
    \sum_{\tup{f}^\pm\in\Omega(\tup{F})}\langle \Psi | \psi(\tup{f}^+)\rangle Q^{\tup{F}|p}_{\tup{t}|t_0}(\tup{f}^+,\tup{f}^-)\langle \psi(\tup{f}^-)|\Phi\rangle
    \quad\text{for any $|\Psi\rangle,|\Phi\rangle\in \mathbb{C}^d$,}
\end{align}
is an inner product in $\mathbb{C}^d$, and thus, it can be used to define a Hilbert space\footnote{
    In fact, $\langle\!\langle\cdot|\cdot\rangle\!\rangle_{\tup{t}|t_0}^{\tup{F}|p}$ is ``almost'' an inner product. At this point it is not guaranteed that $\langle\!\langle \Psi|\Psi\rangle\!\rangle^{\tup{F}|p}_{\tup{t}|t_0} = 0$ only when $|\Psi\rangle$ is a zero vector; in other words, it is possible that $\operatorname{rank}(\hat Q^{\bm F_n|p}_{\tup{t}|t_0}) < d$. This can be easily fixed by collecting all nullified vectors into the set $N = \big\{ |\Psi\rangle\in\mathbb{C}^{d}\mid \langle\!\langle \Psi|\Psi\rangle\!\rangle_{\tup{t}|t_0}^{\tup{F}|p} = 0\big\}$ and switching to the quotient vector space $V = \mathbb{C}^{d}/N$, with elements in the form of equivalence classes $|[\Psi]\rangle = |\Psi\rangle + N$. Then, $\operatorname{rank}(\hat Q^{\bm F_n|p}_{\tup{t}|t_0}) = \operatorname{dim}(V)$ and the function on $V\times V$ defined by $\langle\!\langle [\Psi] | [\Phi]\rangle\!\rangle^{\tup{F}|p}_{\tup{t}|t_0} := \langle\!\langle \Psi | \Phi \rangle\!\rangle^{\tup{F}|p}_{\tup{t}|t_0}$ is a proper inner product~\cite{Gudder_MathSlov12}. Henceforth it is assumed that this construction is employed when needed.
}
\begin{align}
    \mathcal{H}_{\tup{t}|t_0}^{\tup{F}|p} = \Big(\mathbb{C}^d\ ,\ 
    \langle\!\langle\,\cdot\,|\,\cdot\,\rangle\!\rangle^{\tup{F}|p}_{\tup{t}|t_0}\Big).
\end{align}
The bi-probability used to define the inner product can then be interpreted as a matrix element of the product's \textit{metric}:
\begin{align}
    \langle\!\langle\Psi|\Phi\rangle\!\rangle^{\tup{F}|p}_{\tup{t}|t_0}
    = \langle\Psi|\Big(\sum_{\tup{f}^\pm}
    |\psi(\tup{f}^+)\rangle Q^{\tup{F}|p}_{\tup{t}|t_0}(\tup{f}^+,\tup{f}^-)\langle \psi(\tup{f}^-)|\Big)|\Phi\rangle
    \equiv \langle\Psi|\hat Q^{\tup{F}|p}_{\tup{t}|t_0}|\Phi\rangle,
\end{align}
such that
\begin{align}
    \hat Q^{\tup{F}|p}_{\tup{t}|t_0} \geq 0\quad
    \text{and}\quad
    \operatorname{tr}\big(\hat Q^{\tup{F}|p}_{\tup{t}|t_0}\big) =
    \sum_{\tup{f}}\langle\psi(\tup{f})|\hat Q^{\tup{F}|p}_{\tup{t}|t_0}|\psi(\tup{f})\rangle
    =\sum_{\tup{f}}Q^{\tup{F}|p}_{\tup{t}|t_0}(\tup{f},\tup{f}) = 1.
\end{align}
Thus, we arrive at the inner-product representation of bi-probability distributions:
\begin{align}\label{eq:inner_rep}
    Q^{\tup{F}|p}_{\tup{t}|t_0}(\tup{f}^+,\tup{f}^-) &=
    \langle\!\langle \psi(\tup{f}^+)|\psi(\tup{f}^-)\rangle\!\rangle^{\tup{F}|p}_{\tup{t}|t_0}
    = \langle\psi(\tup{f}^+)|\,
    \hat Q^{\tup{F}|p}_{\tup{t}|t_0}\,|\psi(\tup{f}^-)\rangle.
\end{align}

This mathematical result shows an inherent affinity between bi-probabilities and Hilbert spaces. Seemingly, it is the simplest explanation why the formalism of quantum theory is based on the structures of those spaces, subject to constraints imposed by the phenomenology of sequential measurements.

\subsection{The measurement--projector link}\label{sec:obs--projector_link}

As a first step after the establishment of the connection between bi-probabilities and Hilbert spaces, we wish to deduce the measurement--projector link. In the standard formalism, the link is thought to relate the physical action of the measuring device with the action of a projector operator as a formal representation of that device. Here, we begin with the inner-product representation~\eqref{eq:inner_rep} applied in the simplest case of a single-time bi-probability associated with a perfectly fine-grained $K$-device,
\begin{align}
    Q_{t|t_0}^{K|p}(k;k') & =\delta_{k,k'} P^{K|p}_{t|t_0}(k),
\end{align}
which results in the following:
\begin{align}
\nonumber
    Q^{K|p}_{t|t_0}(k;k') &= \delta_{k,k'}\langle\!\langle \psi_{t|t_0}^{K|p}(k)|\psi^{K|p}_{t|t_0}(k)\rangle\!\rangle^{K|p}_{t|t_0} =\delta_{k,k'}\big\langle\psi_{t|t_0}^{K|p}(k)\big|\,\hat Q^{K|p}_{t|t_0}\,\big|\psi_{t|t_0}^{K|p}(k)\big\rangle\\
\nonumber
    &= \delta_{k,k'}\operatorname{tr}_{\mathcal{H}(K)}\left(\,\big|\psi^{K|p}_{t|t_0}(k)\big\rangle\big\langle \psi^{K|p}_{t|t_0}(k)\big|\,\hat Q^{K|p}_{t|t_0}\right)\\
    &\equiv \delta_{k,k'}\operatorname{tr}_{\mathcal{H}(K)}\left(\hat\pi^{K|p}_{t|t_0}\hat Q^{K|p}_{t|t_0}\right).
\end{align}
Here, $(\mathbb{C}^{|\Omega(K)|},\langle\!\langle\cdot|\cdot\rangle\!\rangle^{K|p}_{t|t_0})$ is the Hilbert space equipped with the inner product induced by the bi-probability and $\mathcal{H}(K) = (\mathbb{C}^{|\Omega(K)|},\langle\cdot|\cdot\rangle)$ denotes the Hilbert space corresponding to the same vector space but endowed with the standard inner product. The basis $\{|\psi^{K|p}_{t|t_0}(k)\rangle\in \mathbb{C}^{|\Omega(K)|}\mid k\in\Omega(K)\}$ is chosen in such a way that its elements are mutually orthogonal in $\mathcal{H}(K)$, $\langle \psi^{K|p}_{t|t_0}(k)|\psi^{K|p}_{t|t_0}(k')\rangle = \delta_{k,k'}$; consequently, the operators $\hat\pi^{K|p}_{t|t_0}=|\psi^{K|p}_{t|t_0}(k)\rangle\langle \psi^{K|p}_{t|t_0}(k)|$ are the standard rank-$1$ projectors, and the metric satisfies the standard conditions: $\hat Q^{K|p}_{t|t_0}\geq 0$ and $\operatorname{tr}_{\mathcal{H}(K)}\hat Q^{K|p}_{t|t_0} = 1$.

When we repeat the construction for some coarse-grained variant of the $K$-device---an $\fbar K$-device---we obtain analogous results,
\begin{align}
    Q_{t|t_0}^{\fbar K|p}(\fbar k;\fbar{k}{}') &= \delta_{\fbar k,\fbar{k}{}'}\langle\!\langle\phi^{\fbar K|p}_{t|t_0}(\fbar k)|\phi^{\fbar K|p}_{t|t_0}(\fbar{k})\rangle\!\rangle^{\fbar K|p}_{t|t_0} 
        = \delta_{\fbar k,\fbar{k}{}'}\operatorname{tr}_{\mathcal{H}(\fbar K)}\left(\big|\phi_{t|t_0}^{\fbar K|p}(\fbar k)\big\rangle\big\langle\phi^{\fbar K|p}_{t|t_0}(\fbar{k})\big|\hat Q^{\fbar K|p}_{t|t_0}\right),
\end{align}
where, again, $\mathcal{H}(\fbar K) = (\mathbb{C}^{|\Omega(\fbar K)|},\langle\cdot|\cdot\rangle)$ is the Hilbert space obtained by endowing with the standard inner product the vector space $\mathbb{C}^{|\Omega(\fbar K)|}$. This space is spanned by an orthonormal basis $\{ |\phi^{\fbar K|p}_{t|t_0}(\fbar k)\rangle\in \mathbb{C}^{|\Omega(\fbar K)|}\mid \fbar k\in\Omega(\fbar K)\}$, so that $|\phi^{\fbar K|p}_{t|t_0}(\fbar k)\rangle\langle\phi^{\fbar K|p}_{t|t_0}(\fbar k)|$ are rank-$1$ projector operators, and $\hat Q^{\fbar K|p}_{t|t_0}\geq 0$, $\operatorname{tr}_{\mathcal{H}(\fbar K)}\hat Q^{\fbar K|p}_{t|t_0} = 1$.

The measurements with $K$- and $\fbar K$-devices, of course, adhere to the phenomenological observation~\eqref{obs:coarse-grained_measurement},
\begin{align}
    Q^{\fbar{K}|p}_{t|t_0}(\fbar{k};\fbar k) = P^{\fbar K|p}_{t|t_0}(\fbar k) &= \sum_{k\in\omega(\fbar k)}P_{t|t_0}^{K|p}(k) = \sum_{k\in\omega(\fbar k)} Q_{t|t_0}^{K|p}(k;k),
\end{align}
which then allows us to establish the relation between their respective representations,
\begin{align}\label{eq:low_res_law}
    \operatorname{tr}_{\mathcal{H}(\fbar K)}\Big(\big|\phi^{\fbar K|p}_{t|t_0}(\fbar k)\big\rangle\big\langle\phi_{t|t_0}^{\fbar K|p}(\fbar{k})\big|\,\hat Q^{\fbar K|p}_{t|t_0}\Big) 
    &=\operatorname{tr}_{\mathcal{H}(K)}\Big(\sum_{k\in\omega(\fbar k)}\big|\psi^{K|p}_{t|t_0}(k)\big\rangle\big\langle\psi^{K|p}_{t|t_0}(k)\big|\,\hat Q^{K|p}_{t|t_0}\Big).
\end{align}

Now, there is no purely logical reason for the Hilbert spaces $\mathcal{H}(K)$ and $\mathcal{H}(\fbar K)$ to be related to each other in any way, especially given the fact that $\dim\mathcal{H}(\fbar K)=|\Omega(\fbar K)|\leq\dim\mathcal{H}(K)=|\Omega(K)|$. However, it seems grossly unnecessary that the respective formal representations ``live'' in different Hilbert spaces, even though, ostensibly, both describe a measurement of the same physical quantity (only with different degree of graining). Therefore, if we wish to be efficient with the ``economy of entities'' in our deduction, we ought to bring both representations to the same higher-dimensional space $\mathcal{H}(K)$. 

To accomplish this, first we have to resolve a conundrum: we have to decide how to extend the projectors $|\phi^{\fbar K|p}_{t|t_0}(\fbar k)\rangle\langle\phi_{t|t_0}^{\fbar K|p}(\fbar{k})|$ and the metric $\hat Q^{\fbar K|p}_{t|t_0}$ from the lower-dimensional space $\mathcal{H}(\fbar K)$ to the higher-dimensional $\mathcal{H}(K)$.
Generally, there are infinitely many ways to do this, but here we can rely on the phenomenological relation~\eqref{eq:low_res_law} as a guiding principle for our decision making. The simplest choice adhering to this principle is to identify the operators in $\mathcal{H}(\fbar K)$ with operators in $\mathcal{H}(K)$ that fulfill the same function, i.e., let projectors correspond to projectors, and metrics correspond to metrics,
\begin{align}
    \big|\phi^{\fbar K|p}_{t|t_0}(\fbar k)\big\rangle\big\langle\phi^{\fbar K|p}_{t|t_0}(\fbar{k})\big| \to 
        \hat \pi_{t|t_0}^{\fbar K|p}(\fbar k) := \sum_{k\in\omega(\fbar k)}\hat\pi^{K|p}_{t|t_0}(k)
        \quad\text{and}\quad \hat Q^{\fbar K|p}_{t|t_0} \to \hat Q^{K|p}_{t|t_0}.
\end{align}

Next, let us consider a fine-grained observable $L$ measured with an $L$-device that is \textit{inequivalent} to the $K$-device. In such a case, again we arrive at the corresponding parameterization with the basis $\{|\chi^{L|p}_{t|t_0}(\ell)\rangle\in \mathbb{C}^{|\Omega(L)|}\mid \ell\in\Omega(L)\}$ spanning the Hilbert space $\mathcal{H}(L)=(\mathbb{C}^{|\Omega(L)|},\langle\cdot|\cdot\rangle)$, rank-$1$ projectors $|\chi^{L|p}_{t|t_0}(\ell)\rangle\langle\chi^{L|p}_{t|t_0}(\ell)|$, and the metric $\hat Q^{L|p}_{t|t_0}$. Since the measurements are still performed on the same physical system, we should also extend this representation to the first Hilbert space $\mathcal H(K)$ following the same principle as with the $\fbar K$-device: the projectors in $\mathcal{H}(L)$ are mapped to rank-$1$ projectors in $\mathcal H(K)$, $|\chi^{L|p}_{t|t_0}(\ell)\rangle\langle\chi^{L|p}_{t|t_0}(\ell)|\to\hat \pi^{L|p}_{t|t_0}(\ell)$ (where $\hat \pi_{t|t_0}^{L|p}(\ell)\hat \pi_{t|t_0}^{L|p}(\ell') = \delta_{\ell,\ell'}\hat \pi_{t|t_0}^{L|p}(\ell)$, $\sum_{\ell\in\Omega(L)}\hat \pi^{L|p}_{t|t_0}(\ell) = \hat 1$), and the metric is mapped to the original operator in $\mathcal{H}(K)$, $\hat Q^{L|p}_{t|t_0} \to \hat Q^{K|p}_{t|t_0}$. The coarse-grained variants of $L$ should then be treated in an analogous manner as $\fbar K$: the metric would remain fixed and $\hat\pi^{\fbar L|p}_{t|t_0}(\fbar\ell) = \sum_{\ell\in\omega(\fbar\ell)}\hat\pi^{L|p}_{t|t_0}(\ell)$.

Of course, just like it was with $\fbar K$, the mappings for $L$ also must be consistent with the phenomenology,
\begin{align}\label{eq:sys_space_extension_X}
    P^{L|p}_{t|t_0}(\ell) = \operatorname{tr}_{\mathcal{H}(L)}\left(|\chi^{L|p}_{t|t_0}(\ell)\rangle\langle\chi^{L|p}_{t|t_0}(\ell)|\,\hat Q^{L|p}_{t|t_0}\right) = \operatorname{tr}_{\mathcal{H}(K)}\left(\hat \pi^{L|p}_{t|t_0}(\ell)\,\hat Q^{K|p}_{t|t_0}\right),
\end{align}
which implies that the set of projectors representing the measurement with the $L$-device in $\mathcal{H}(K)$ space \textit{cannot} be the same projectors representing the action of the $K$-device, $\{\hat\pi^{K|p}_{t|t_0}(k)\mid k\in\Omega(k)\}\neq\{\hat\pi^{L|p}_{t|t_0}(\ell)\mid \ell\in\Omega(L)\}$. If this were not the case, then the probability distributions of both devices would be identical, thus contradicting the assumption that $L$ and $K$ are inequivalent observables.

By following the guiding principle that for every system observable $F$---be it perfectly fine-grained, or a coarse-grained variant of some fine-grained observable---we map the corresponding metric $\hat Q^{F|p}_{t|t_0}$ onto the original operator $\hat Q^{K|p}_{t|t_0}$, we have forced it into the role of the Hilbert space representation of the initialization event. Indeed, since the same metric is selected for every measurement, this operator becomes independent of the choice of measuring device, and by extension, also of the timing of the device deployment; hence, for every $F$-device (including the original $K$-device) and $t>t_0$:
\begin{align}
    \hat Q^{F|p}_{t|t_0} \equiv \hat\rho^{\,p}_{t_0},\quad\text{such that $\hat\rho_{t_0}^{\,p}\geq 0$ and $\operatorname{tr}\hat\rho^{\,p}_{t_0} = 1$.}    
\end{align}
On the other hand, as the measurement is carried out by the observer, who has the full autonomy in choosing the device and its deployment time, the projectors ought to be independent of the initialization event; and thus, for every $f\in\Omega(F)$, $t>t_0$, and distribution $p$:
\begin{align}
\begin{array}{r}
\textbf{Hilbert space}\\
\textbf{representation}\\
\text{(Measurement-}\\
\text{projector link)}
\end{array}\quad
    \hat\pi^{F|p}_{t|t_0}(f) \equiv \hat P^F_t(f)\ :\ \hat P^F_t(f)\hat P^F_t(f')=\delta_{f,f'}\hat P^F_t(f),\ \sum_{f\in\Omega(F)}\hat P^F_t(f) = \hat 1.
\end{align}
In this way, we have arrived naturally at the link between measuring devices and partitions of the Hilbert space into images of the set of orthogonal projectors representing the device.

Finally, as we have managed to confine all elements of the representation to one Hilbert space, at this point it is no longer necessary to label the space with the observable $K$, which was chosen arbitrarily in the first place. Rather, the Hilbert space that houses the projectors and the metric---which together form the representation of single-time bi-probabilities---can be considered as a property of the measured system itself; hence, we can now speak of the Hilbert space $\mathcal{H}_S$ that characterizes a given quantum system $S$.

\subsection{Hilbert space representation for initialization events}\label{sec:initialization_event}
When discussing the phenomenology of the Markov property (observation~\ref{obs:markov}) in Part I we had determined that the initialization is accomplished via the deployment of a perfectly fine-grained measuring device, see Eq.~\eqref{eq:phenomenology_of_initialization}. We can now use this insight, together with the measurement--projector link just established, to deduce the specific form of the metric $\hat\rho_{t_0}^{\,p}$ representing a given initialization event described phenomenologically by the distribution $p$. To this end, first consider an experiment in which the perfectly fine-grained $K$-device is deployed immediately after the initialization with the same device measuring result $k_0$, i.e., $p^{K'}(k') = \delta_{K',K}\delta_{k',k_0} \equiv \delta^K_{k_0}(K',k')$. In accordance with the phenomenological observations~\ref{obs:markov} and~\ref{obs:uncertainty}, the results of such a measurement are described by
\begin{align}
   \lim_{\Delta t \to 0^+} P^{K|\delta^K_{k_0}}_{t_0+\Delta t|t_0}(k) &= \lim_{\Delta t\to 0} P^{K|K}_{t_0+\Delta t|t_0}(k|k_0) = C^{K|K}_{k,k_0} = \delta_{k,k_0}.
\end{align}
On the other hand, by using the Hilbert space representation of the measurement we have
\begin{align}
    \lim_{\Delta t\to 0^+}P^{K|\delta^K_{k_0}}_{t_0+\Delta t|t_0}(k) &= \operatorname{tr}_S\big[
        \hat P^{K}_{t_0}(k)\hat\rho_{t_0}^{\delta^K_{k_0}}\big] 
        = \langle\Psi^{K}_{t_0}(k)|\hat\rho^{\delta^K_{k_0}}_{t_0}|\Psi^{K}_{t_0}(k)\rangle,
\end{align}
where $\{|\Psi^{K}_{t_0}(k)\rangle\in\mathcal{H}_S\mid k\in\Omega(K)\}$ form an orthonormal basis (the projectors representing perfectly fine-grained measurements are all rank-$1$). This, together with the requirements $\hat\rho_{t_0}^{\,p}\geq0$ and $\operatorname{tr}_S\hat\rho^{\,p}_{t_0} = 1$ for any $p^K(k)$, implies that the metric matrix representing the initialization event in the form of a fine-grained measurement is itself a rank-$1$ projection operator,
\begin{align}
\begin{array}{r}
    \textbf{Hilbert space representation}\\
    \text{(Initialization event)}
\end{array}\quad
    \hat\rho_{t_0}^{\delta^K_{k_0}} = \hat P^{K}_{t_0}(k_0) = |\Psi^{K}_{t_0}(k_0)\rangle\langle \Psi^{K}_{t_0}(k_0)|.
\end{align}
And so, the metric representing a general initialization event has the form of a convex combination of rank-$1$ projectors corresponding to the perfectly fine grained measurements,
\begin{align}
    \hat\rho_{t_0}^{\,p} = \sum_{K}\sum_{k\in\Omega(K)}p^K(k) \hat \rho^{\delta^K_{k}}_{t_0} =\sum_{K}\sum_{k\in\Omega(K)}p^K(k) \hat P^{K}_{t_0}(k)
        = \sum_{K}\sum_{k\in\Omega(K)}p^K(k) |\Psi^{K}_{t_0}(k)\rangle\langle\Psi^{K}_{t_0}(k)|.
\end{align}

We have successfully deduced that a (single) measurement with a given device formally corresponds to a set of orthogonal projectors. Then, it stands to reason that the reverse is also true: in principle, it should be possible to build a device such that its measurements can be represented by a given set of projectors. Bearing this observation in mind, consider the following formal statement:
\begin{align}
    P^{K|\delta^K_{k_0}}_{t|0}(k) &= \operatorname{tr}_S\big[\hat P_{t}^{K}(k)\hat P^{K}_{0}(k_0)]
        \equiv \operatorname{tr}_S\big[\hat P^{K_t}_{0}(k)\hat P^{K}_{0}(k_0)\big]
            = \lim_{\Delta t\to 0^+}P^{K_t|K}_{\Delta t|0}(k|k_0) = C_{k,k_0}^{K_t|K}.
\end{align}
Here, $K_t$ denotes an observable such that the projectors corresponding to the deployment of the $K_t$-device at time $t=0$ are the same as the projectors corresponding to the $K$-device deployed at time $t > 0$, i.e., $\hat P_0^{K_t}(k) = \hat P_t^K(k)$---according to our previous observation, such a device should exist in principle. 

The interesting thing about the above ``trick'' is that we have exploited the formal measurement--projector link to essentially eliminate the passage of time from the picture. We have shown here that, in the context of single measurement experiments, the effects of the system dynamics unfolding during the time between the initialization and the terminating measurement are fully captured by the measurement with an appropriate device immediately following the initialization. In this sense, the single-measurement context can be considered as \textit{quantum statics}, where the metric $\hat\rho_{t_0}^{\,p}$ is \textit{de facto} the master object of the formalism (i.e., all possible outcomes are derivable from it), and all phenomenologically answerable questions can be expressed in terms of a deployment of a certain measuring device in the instant following the initialization event. 

Soon it will become apparent that the static picture above does not hold anymore in contexts of sequential measurements, in which the dynamical laws of the system can no longer be simply ``absorbed'' by the freedom of choice of the measuring devices. We will further investigate the formal relations between representations of $K_t$ and $K$ in the upcoming Section~\ref{sec:dynamical_laws}, where we shall deduce the Hilbert space representation of the system dynamical laws.

\subsection{Hilbert space representation for bi-probabilities}\label{sec:H-rep_for_bi-probs}

Consider a conditional probability distribution,
\begin{align}\label{eq:n=1_conditional_prob}
    P^{F_{n+1}|\tup{F}}_{t_{n+1}|\bm{t}_{n}}(f_{n+1}|\bm{f}_{n}) &= \frac{P^{\bm{F}_{n+1}|\rho}_{\bm{t}_{n+1}|t_0}(\bm{f}_{n+1})}{P^{\tup{F}|\rho}_{\bm{t}_{n}|t_0}(\bm{f}_{n})}
        = \frac{Q^{\bm{F}_{n+1}|\rho}_{\bm{t}_{n+1}|t_0}(\bm{f}_{n+1},\bm{f}_{n+1})}{Q^{\tup{F}|\rho}_{\bm{t}_{n}|t_0}(\bm{f}_{n},\bm{f}_{n})}.
\end{align}
Formally, this is an $n=1$ bi-probability, and so we can construct its system Hilbert space representation in the same fashion as it was previously done in Section~\ref{sec:obs--projector_link},
\begin{align}
    P^{F_{n+1}|\bm{F}_{n}}_{t_{n+1}|\bm{t}_{n}}(f_{n+1}|\bm{f}_{n}) &= \operatorname{tr}_{S}\left[\hat P^{F_{n+1}}_{t_{n+1}}(f_{n+1})
        \frac{\hat q^{\tup{F}|\rho}_{\bm{t}_{n}|t_0}(\bm{f}_{n})}{Q^{\tup{F}|\rho}_{\bm{t}_{n}|t_0}(\bm{f}_{n},\bm{f}_{n})}
    \right],
\end{align}
where we have introduced the operator $\hat q^{\tup{F}|\rho}_{\bm{t}_{n}|t_0}(\bm{f}_{n})\geqslant 0$ that plays the role of the (non-normalized) metric. Then, using the completeness of the projector operators, $\sum_{f\in\Omega(F)}\hat P^{F}_t(f) = \hat 1$ for any $F$, we find out that the bi-probability is given by the trace of the pseudo-metric $\hat q^{\tup{F}|\rho}_{\tup{t}|t_0}$,
\begin{align}
\nonumber
    Q_{\tup{t}|t_0}^{\tup{F}|\rho}(\tup{f},\tup{f}) 
    &= \!\!\sum_{f_{n+1}\in\Omega(F_{n+1})}\!\!\!\!
    P^{F_{n+1}|\tup{F}}_{t_{n+1}|\tup{t}}(f_{n+1}\mid\tup{f})Q_{\tup{t}|t_0}^{\tup{F}|\rho}(\tup{f},\tup{f})
    = \operatorname{tr}_S\Big[\sum_{f_{n+1}}\hat P^{F_{n+1}}_{t_{n+1}}(f_{n+1})\hat q^{\tup{F}|\rho}_{\tup{t}|t_0}(\tup{f})\Big]\\
    &= \operatorname{tr}_S\big[\hat q^{\tup{F}|\rho}_{\tup{t}|t_0}(\tup{f})\big],
\end{align}
and thus,
\begin{align}
    \operatorname{tr}_{S}\left[\hat q^{\bm{F}_{n+1}|\rho}_{\bm{t}_{n+1}|t_0}(\bm{f}_{n+1})\right] &= \operatorname{tr}_{S}\left[
        \hat P^{F_{n+1}}_{t_{n+1}}(f_{n+1})\hat q^{\tup{F}|\rho}_{\bm{t}_{n}|t_0}(\bm{f}_{n})
    \right]
    = \operatorname{tr}_{S}\left[
        \hat P^{F_{n+1}}_{t_{n+1}}(f_{n+1})\hat q^{\tup{F}|\rho}_{\bm{t}_{n}|t_0}(\bm{f}_{n})\hat P^{F_{n+1}}_{t_{n+1}}(f_{n+1})
    \right].
\end{align}
This recurrence equation, with the initial condition
\begin{align}
    \hat q^{F_1|\rho}_{t_1|t_0}(f_1) = \hat P_{t_1}^{F_1}(f_1)\hat\rho_{t_0}\hat P_{t_1}^{F_1}(f_1);
    \quad
    \hat\rho_{t_0} = \sum_{K,k}\rho^K(k)\hat P^K_{t_0}(k);
    \ 
    \sum_{K,k}\rho^K(k)=1, \rho^K(k)\geq 0;
\end{align}
admits the following solution:
\begin{align}
    \hat q^{\tup{F}|\rho}_{\tup{t}|t_0}(\tup{f}) &= \hat P^{F_n}_{t_n}(f_n)\cdots\hat P^{F_1}_{t_1}(f_1)\hat\rho_{t_0}\hat P^{F_1}_{t_1}(f_1)\cdots\hat P^{F_n}_{t_n}(f_n) \geqslant 0,
\end{align}
which then leads to the parameterization of the diagonal part of the bi-probability,
\begin{align}
    Q^{\tup{F}|\rho}_{\tup{t}|t_0}(\tup{f},\tup{f}) = P^{\tup{F}|\rho}_{\tup{t}|t_0}(\tup{f}) = \operatorname{tr}_{S}\Big[
        \hat P^{F_n}_{t_n}(f_n)\cdots \hat P^{F_1}_{t_1}(f_1)\hat\rho_{t_0}\hat P^{F_1}_{t_1}(f_1)\cdots\hat P^{F_n}_{t_n}(f_n)
    \Big].
\end{align}

We can generalize this parameterization to the off-diagonal of $Q_{\tup{t}|t_0}^{\tup{F}|\rho}$ by making the use of the bi-consistency relation~\ref{prop:bi-prob:bi-consistency}:
% \begin{align}
% \nonumber
%     &\sum_{f_j^+\neq f_j^-}Q^{\tup{F}|\rho}_{\tup{t}|t_0}(
%             f_n,\ldots,f_j^+,\ldots,f_1\,;\,
%             f_n,\ldots,f_j^-,\ldots,f_1
%         )\\
% \label{eq:bi-consistency}
%     &\phantom{=}=Q^{
%         F_n,\ldots,\cancel{F_j},\ldots,F_1|\rho
%     }_{
%         t_n,\ldots,\cancel{t_j},\ldots,t_1|t_0
%     }(
%         f_n,\ldots,\cancel{f_j},\ldots,f_1\,;\,
%         f_n,\ldots,\cancel{f_j},\ldots,f_1
%     ) - \sum_{f_j\in\Omega(F_j)}Q^{\tup{F}|\rho}_{\tup{t}|t_0}(\tup{f}, \tup{f}).
% \end{align}
\begin{align}
    \nonumber
    &Q^{
        F_n,\ldots,\cancel{F_j},\ldots,F_1|\rho
    }_{
        t_n,\ldots,\cancel{t_j},\ldots,t_1|t_0
    }(
        f_n,\ldots,\cancel{f_j},\ldots,f_1\,;\,
        f_n,\ldots,\cancel{f_j},\ldots,f_1
    )
    \\\nonumber
    &\phantom{=}
    = \sum_{f_j^\pm\in\Omega(F_j)}
        Q^{\tup{F}|\rho}_{\tup{t}|t_0}(\tup{f}^+,\tup{f}^-)
    =\Bigl\{
        \sum_{f^+_j = f_j^-}+\sum_{f_j^+\neq f_j^-}\Bigr\}\,
        Q^{\tup{F}|\rho}_{\tup{t}|t_0}(\tup{f}^+,\tup{f}^-)
    \\\nonumber
    \implies\ &
    \sum_{f_j^+\neq f_j^-}Q^{\tup{F}|\rho}_{\tup{t}|t_0}(
            f_n,\ldots,f_j^+,\ldots,f_1\,;\,
            f_n,\ldots,f_j^-,\ldots,f_1
        )
    \\\label{eq:bi-consistency}
    &\phantom{=}=Q^{
        F_n,\ldots,\cancel{F_j},\ldots,F_1|\rho
    }_{
        t_n,\ldots,\cancel{t_j},\ldots,t_1|t_0
    }(
        f_n,\ldots,\cancel{f_j},\ldots,f_1\,;\,
        f_n,\ldots,\cancel{f_j},\ldots,f_1
    ) - \sum_{f_j\in\Omega(F_j)}Q^{\tup{F}|\rho}_{\tup{t}|t_0}(\tup{f}, \tup{f}).
\end{align}
First, we have
\begin{align}
\nonumber
    &Q^{
        F_n,\ldots,\cancel{F_j},\ldots,F_1|\rho
    }_{
        t_n,\ldots,\cancel{t_j},\ldots,t_1|t_0
    }(
        f_n,\ldots,\cancel{f_j},\ldots,f_1\,;\,
        f_n,\ldots,\cancel{f_j},\ldots,f_1)\\
\nonumber
    &\phantom{=}=P^{
        F_n,\ldots,\cancel{F_j},\ldots,F_1|\rho
    }_{
        t_n,\ldots,\cancel{t_j},\ldots,t_1|t_0
    }(
        f_n,\ldots,\cancel{f_j},\ldots,f_1
    )\\
\nonumber
    &\phantom{=}
    = \operatorname{tr}_S\Big[\hat P^{F_n}_{t_n}(f_n)\cdots\hat P^{F_{j+1}}_{t_{j+1}}(f_{j+1})\Big(\sum_{f_j^+\in\Omega(F_j)}\hat P^{F_j}_{t_j}(f_j^+)\Big)\hat P^{F_{j-1}}_{t_{j-1}}(f_{j-1})\cdots\hat P^{F_1}_{t_1}(f_1)\hat\rho_{t_0}\\
    &\phantom{==\operatorname{tr}_S}\times
       \hat P^{F_1}_{t_1}(f_1)\cdots\hat P^{F_{j-1}}_{t_{j-1}}(f_{j-1})\Big(\sum_{f_j^-\in\Omega(F_j)}\hat P^{F_j}_{t_j}(f_j^-)\Big)\hat P^{F_{j+1}}_{t_{j+1}}(f_{j+1})\cdots\hat P^{F_n}_{t_n}(f_n)
    \Big],
\end{align}
and so Eq.~\eqref{eq:bi-consistency} can be rewritten as
\begin{align}
\nonumber
    &\sum_{f_j^+\neq f_j^-} Q^{\tup{F}|\rho}_{\tup{t}|t_0}(f_n,\ldots,f_j^+,\ldots,f_1;f_n,\ldots,f_j^-,\ldots,f_1)\\
\nonumber
    &\phantom{=}=
        P^{F_n,\ldots,\cancel{F_j},\ldots,F_1|\rho}_{t_n,\ldots,\cancel{t_j},\ldots,t_1|t_0}(f_n,\ldots,\cancel{f_j},\ldots,f_1) - \sum_{f_j\in\Omega(F_j)}P_{\tup{t}|t_0}^{\tup{F}|\rho}(\tup{f})\\
\nonumber
    &\phantom{=}=
        \sum_{f_j^\pm\in\Omega(F_j)}\!\!\!\operatorname{tr}_S\Big[
            \hat P^{F_n}_{t_n}(f_n)\cdots \hat P^{F_j}_{t_j}(f_j^+)\cdots\hat P_{t_1}^{F_1}(f_1)\hat\rho_{t_0}\hat P_{t_1}^{F_1}(f_1)
            \cdots\big(\hat P^{F_j}_{t_j}(f_j^-)-\hat P^{F_j}_{t_j}(f_j^+)\big)\cdots\hat P_{t_n}^{F_1}(f_n)
        \Big]\\
    &\phantom{=}=
        \sum_{f_j^+\neq f_j^-}\operatorname{tr}_S\Big[
            \hat P^{F_n}_{t_n}(f_n)\cdots\hat P^{F_j}_{t_j}(f_j^+)\cdots\hat P^{F_1}_{t_1}(f_1)\hat\rho_{t_0}
            \hat P^{F_1}_{t_1}(f_1)\cdots\hat P^{F_j}_{t_j}(f_j^-)\cdots\hat P^{F_n}_{t_n}(f_n)
        \Big].
\end{align}
Since the inconsistency relation holds for all $n\in\mathbb{N}$, $1\leq j \leq n$, $f_j\in\Omega(F_j)$, $t_n > \cdots > t_1 > t_0 > 0$, we arrive at the following conclusion:
\begin{align}\label{prop:bi-prob:H-space_rep_multi-obs}
    \begin{array}{r}
    \textbf{Hilbert space}\\
    \textbf{representation}\\
    \text{(Bi-probability}\\
    \text{distributions)}\\
    \end{array}\quad 
    Q_{\tup{t}|t_0}^{\tup{F}|\rho}(\tup{f}^+,\tup{f}^-) &= \operatorname{tr}_{S}\Big[
        \Big(\prod_{j=n}^1\hat P^{F_j}_{t_j}(f_j^+)\Big)\hat\rho_{t_0}
        \Big(\prod_{j=1}^n\hat P_{t_j}^{F_j}(f_j^-)\Big)
    \Big],
\end{align}
where $\prod_{j=n}^1\hat A_j$ ($\prod_{j=1}^n\hat A_j$) denotes an ordered product of operators $\hat A_n\cdots\hat A_1$ ($\hat A_1\cdots\hat A_n$), the metric matrix $\hat\rho_{t_0}$ is given by
\begin{align*}
    \hat\rho_{t_0} = \sum_{K}\sum_{k\in\Omega(K)}\rho^K(k)\hat P_{t_0}^{K}(k) 
        = \sum_{K}\sum_{k\in\Omega(K)}\rho^K(k)|\Psi^K_{t_0}(k)\rangle\langle\Psi^K_{t_0}(k)|,
\end{align*}
and it represents the initialization event defined by the phenomenological distribution $\rho^K(k)$ referring to measurements done with perfectly fine-grained $K$-devices deployed at time $t_0$.

This result demonstrates the limitations of the measurement--projector link: the alluring idea that ``the action of the $F$-device at time $t$ is in one-to-one correspondence with the action of the projector operator $\hat P_t^F(f)$'', which works for a single measurement, does not extend to sequential measurements. To see this, notice that, in the formal representation of a probability distribution for a sequence of measurements,
\begin{align}
    P^{\tup{F}|\rho}_{\tup{t}|t_0}(\tup{f}) 
    &= Q^{\tup{F}|\rho}_{\tup{t}|t_0}(\tup{f},\tup{f}) =
    \operatorname{tr}_S\Big[
        \Big(\prod_{j=n}^1\hat P^{F_j}_{t_j}(f_j)\Big)^\dagger\Big(\prod_{j=n}^1\hat P^{F_j}_{t_j}(f_j)\Big)\hat\rho_{t_0}
    \Big]\equiv \operatorname{tr}_S\Big[\hat E_{\tup{t}}^{\tup{F}}(\tup{f})\hat\rho_{t_0} \Big],
\end{align}
the operator $\hat E^{\tup{F}}_{\tup{t}}(\tup{f})$ is generally \textit{not} a projector. Therefore, given that $\hat\rho_{t_0}$ represents the initialization event, we cannot say that the action of a sequence of measuring devices is represented in an analogous way as the action of a single measuring device. Moreover, it is not true that the action of a sequence of devices [represented by $\hat E_{t_n,\ldots,t_1}^{F_n,\ldots,F_1}(f_n,\ldots,f_1)$] corresponds to a sequence of actions of single devices [represented by $\prod_{j=n}^1\hat P_{t_j}^{F_j}(f_j)$], even though the Hilbert-space representations of both consist of the same elementary building blocks---the projectors $\hat P_{t_j}^{F_j}(f_j)$. This leads us to the following conclusion: since, in general, the results of a sequence of measurements cannot be represented by a deployment of a single measuring device following immediately the initialization event, the quantum statics picture is incapable of describing the context of sequential measurements. Thus, the metric $\hat\rho_{t_0}$ \textit{cannot} be identified as the master object of the formalism in non-static (dynamical) contexts.

\section{Dynamical laws of the system}\label{sec:dynamical_laws}

\subsection{Unitary evolution operator}\label{sec:unitary_evol}

Consider an arbitrarily chosen observable $K$ of the system $S$ that is measured with a perfectly fine-grained $K$-device deployed at time $t = 0$; the projector operators associated with $K$ are all rank-$1$, $\hat P^K_{t=0}(k) = |\Psi^K_0(k)\rangle\langle \Psi^K_0(k)|$, with
\begin{align}
\mathbb{K}_{t=0} = \{|\Psi^K_0(k)\rangle\in \mathcal{H}_S\mid k\in\Omega(K)\}
\end{align}
forming an orthonormal basis in the system's Hilbert space. If the device was deployed at later time $t>0$, the set of operators associated with the measurement would project onto elements of orthonormal basis, which, in most cases, would be different than before, $\mathbb{K}_{t>0}\neq\mathbb{K}_{t=0}$. Then, the way in which the basis $\mathbb{K}_0$ transforms into $\mathbb{K}_{t>0}$ is what we would call the \textit{dynamical law} of the observable measured by $K$-device.

Hence, the passage of time in this representation amounts to the change of basis $\mathbb{K}_0\to\mathbb{K}_t$, and thus, the dynamical law governing the observable's evolution can be represented with an unitary operator-valued function $t\mapsto \hat U_t^K$ executing the corresponding basis transformation:
\begin{align}
    \hat U_t^K := \sum_{k\in\Omega(K)}|\Psi^K_0(k)\rangle\langle\Psi_0^K(k)|\Psi_t^K(k)\rangle\langle\Psi_t^K(k)|.
\end{align}
Then, for example, we can express a single-time probability as a matrix element of the evolution operator:
\begin{align}
    P^{K|K}_{t|0}(k|k_0) &=\operatorname{tr}_S\big[\hat P^K_t(k)\hat P^K_0(k_0)]
    = \operatorname{tr}_S\Big[(\hat U_{t}^K)^\dagger\hat P^K_0(k)\hat U^K_t\hat P_0^K(k_0)]
    = |\langle\Psi_0^K(k)|\hat U_t^K|\Psi^K_0(k_0)\rangle|^2.
\end{align}

Since measurements with any coarse-grained variant of $K$-device are associated with projectors onto subspaces spanned by elements of bases $\mathbb{K}_t$, $\hat U^K_t$ also represents the dynamical law for observables $\fbar K$:
\begin{align}
    \hat P^{\fbar K}_t(\fbar k) &= \sum_{k\in\omega(\fbar k)}\hat P_t^K(k) = \sum_{k\in\omega(\fbar k)}(\hat U_t^K)^\dagger\hat P_0^K(k)\hat U_t^K
        = (\hat U_t^K)^\dagger\Big(\sum_{k\in\omega(\fbar k)}\hat P_0^K(k)\Big)\hat U_t^K = (\hat U_t^K)^\dagger\hat P_0^{\fbar K}(\fbar k)\hat U_t^K.
\end{align}

Moreover, we can conclude from the observation~\ref{obs:uncertainty} that the same operator is also fit to represent dynamical law for any perfectly fine-grained observable inequivalent to $K$: According to the phenomenological uncertainty relations, the correlation function between any fine-grained observables,
\begin{align}
    C^{K|L}_{k,\ell} = \lim_{\Delta t \to 0^+}P^{K|L}_{t+\Delta t|t}(k|\ell) = \lim_{\Delta t\to 0^+}P^{L|K}_{t+\Delta t|t}(\ell|k),
\end{align}
does not depend on the absolute time $t$ when the measuring devices were deployed in a rapid succession; when expressed in terms of its Hilbert-space representation, uncertainty relations translate into the following equation:
\begin{align}
\nonumber
    |\langle\Psi_0^K(k)|\Psi_0^L(\ell)\rangle|^2 &= |\langle\Psi^K_0(k)|\hat U_t^K(\hat U^L_t)^\dagger|\Psi^L_0(\ell)\rangle|^2\\
\label{eq:single_U}
    &= |\langle\Psi^L_0(\ell)|\hat U_t^L(\hat U^K_t)^\dagger|\Psi^K_0(k)\rangle|^2
    = |\langle\Psi^L_0(\ell)|\Psi^K_0(k)\rangle|^2,
\end{align}
where $\mathbb{L}_t = \{ |\Psi^L_t(\ell)\rangle\in \mathcal{H}_S\mid \ell\in\Omega(L)\}\neq\mathbb{K}_t$ is a basis associated with observable $L$ and $\hat U^L_t$ denotes the $\mathbb{L}_t \to \mathbb{L}_0$ transformation. Therefore, the unitary operators representing the dynamical laws for observables $K$ and $L$ must be equal:\footnote{
Formally, Eq.~\eqref{eq:single_U} accepts more general solution: $\hat U^L_t = \exp(-\mathrm{i}\theta_t)\hat U_t^K$; however, the gauge phase $\theta_{t}$ can be set here to zero without loss of generality because it does not influence the transformation of the projectors:
\begin{align*}
    \hat P_t^L(\ell) = (\hat U_t^L)^\dagger\hat P^L_0(\ell)\hat U_t^L = \mathrm{e}^{\mathrm{i}\theta_{t}}(\hat U_t^K)^\dagger\hat P^L_0(\ell)\hat U_t^K \mathrm{e}^{-\mathrm{i}\theta_{t}} = (\hat U_t^K)^\dagger\hat P^L_0(\ell)\hat U_t^K.
\end{align*}
}
\begin{align}
    \hat U_t^L &= \hat U_t^K.
\end{align}

Since the dynamics of each observable is described by the same unitary evolution operator, there is no longer any reason to label it with the observable $K$. In that case, analogously to the Hilbert space of the system $\mathcal{H}_S$, the evolution operator $t\mapsto\hat U_t$ can be ascribed to the quantum system itself as a representation of its intrinsic dynamical law.

\subsection{Generator of the evolution operator}\label{sec:generator_evol}

The short-time behavior of the survival probability implied by the Zeno effect (cf. observation~\ref{obs:zeno}),
\begin{align}\label{eq:surv_short-time}
    P^{K|K}_{t+\Delta t|t}(k|k) &\xrightarrow{\Delta t\to 0^+} 1 - v(k,t)^2\Delta t^2 + O(\Delta t^3),
\end{align}
inevitably constraints the possible forms of time-dependence in the evolution operator $\hat U_t$. In particular, we can immediately conclude that the function $t\mapsto \hat U_t$ has to be continuous and differentiable at each $t$, so that it possesses a well-behaved short-time series expansion,
\begin{align}\label{eq:short-time_Ut}
    \hat U_{t+\Delta t} &= \hat U_t + \Delta t\frac{\mathrm{d}\hat U_t}{\mathrm{d}t} + \frac{\Delta t^2}{2}\frac{\mathrm{d}^2\hat U_t}{\mathrm{d}t^2} + O(\Delta t^3),
\end{align}
which could then produce the corresponding expansion for the probability,
\begin{align}
\nonumber
    P_{t+\Delta t|t}^{K|K}(k|k) &=|\langle \Psi^K_0(k)|\hat U_{t+\Delta t}\hat U_t^\dagger|\Psi^K_0(k)\rangle|^2\\
\nonumber
    &\xrightarrow{\Delta t\to 0^+} 1 
        + \Delta t \langle \Psi^K_0(k)|\Big(\frac{\mathrm{d}\hat U_t}{\mathrm{d}t}\hat U_t^\dagger + \hat U_t\frac{\mathrm{d}\hat U^\dagger_t}{\mathrm{d}t}\Big)|\Psi^K_0(k)\rangle\\
\nonumber
    &\phantom{\xrightarrow{\Delta t\to 0^+} 1}
        + \frac{\Delta t^2}{2}\langle \Psi^K_0(k)|
        \Big(\frac{\mathrm{d}^2\hat U_t}{\mathrm{d}t^2}\hat U_t^\dagger + \hat U_t\frac{\mathrm{d}^2\hat U_t^\dagger}{\mathrm{d}t^2}\Big)
        |\Psi^K_0(k)\rangle\\
    &\phantom{\xrightarrow{\Delta t\to 0^+} 1}
        +\Delta t^2\langle \Psi^K_0(k)|\frac{\mathrm{d}\hat U_t}{\mathrm{d}t}\hat U_t^\dagger|\Psi^K_0(k)\rangle
            \langle \Psi^K_0(k)|\hat U_t\frac{\mathrm{d}\hat U_t^\dagger}{\mathrm{d}t}|\Psi^K_0(k)\rangle + O(\Delta t^3).
\end{align}
In the expansion above, the linear term vanishes and the quadratic term is negative; consequently, these phenomenological inputs constraint the form of $\hat U_t$ even further:
\begin{align}\label{eq:U_t_conditions}
    \frac{\mathrm{d}\hat U_t}{\mathrm{d}t}\hat U_t^\dagger = -\hat U_t\frac{\mathrm{d}\hat U_t^\dagger}{\mathrm{d}t},
    \quad
    \frac{\mathrm{d}^2\hat U_t}{\mathrm{d}t^2}\hat U_t^\dagger = \hat U_t\frac{\mathrm{d}^2\hat U_t^\dagger}{\mathrm{d}t^2},
    \quad
    |\langle\phi|\frac{\mathrm{d}\hat U_t}{\mathrm{d}t}\hat U_t^\dagger|\phi\rangle|^2 \leq -\langle \phi|\frac{\mathrm{d}^2\hat U_t}{\mathrm{d}t^2}\hat U_t^\dagger|\phi\rangle,
\end{align}
for all $|\phi\rangle$ and $t>0$. Bearing these conditions in mind, we can now use the continuity~\eqref{eq:short-time_Ut} of $\hat U_t$ to determine its exact form: given $t>0$, let $s_j = jt/n$, and then
\begin{align}
\nonumber
    \hat U_t 
    &=\lim_{n\to\infty}\hat U_t\prod_{j=n-1}^1 \hat U_{s_j}^\dagger\hat U_{s_j}
    =\lim_{n\to\infty}\hat U_{t}\hat U_{s_{n-1}}^\dagger\Big(\prod_{j=n-2}^1\hat U_{s_j+\frac t n}\hat U_{s_j}^\dagger\Big)
        \hat U_{s_1}
    =\lim_{n\to\infty}\prod_{j=n-1}^0\hat U_{s_j+\frac{t}{n}}\hat U_{s_j}^\dagger\\
\nonumber
    &=\lim_{n\to\infty}
        \prod_{j=n-1}^0\Big(\hat 1 + \frac{t}{n}\frac{\mathrm{d}\hat U_{s_j}}{\mathrm{d}s_j}\,\hat U_{s_j}^\dagger + O(n^{-2}) \Big)
    = \sum_{n=0}^\infty \int_0^t\mathrm{d}s_n\cdots\int_0^{s_2}\mathrm{d}s_1
            \prod_{j=n}^1 \frac{\mathrm{d}\hat U_{s_j}}{\mathrm{d}s_j}\hat U_{s_j}^\dagger\\
    &=\mathcal{T}{\exp}\left[\int_0^t \frac{\mathrm{d}\hat U_{s}}{\mathrm{d}s}\,\hat U_{s}^\dagger\, \mathrm{d}s\right],
\end{align}
and thus, to satisfy all conditions~\eqref{eq:U_t_conditions} and $\hat U_t^\dagger= \hat U_t^{-1}$ for all $t>0$, we must have
\begin{align}
    \frac{\mathrm{d}\hat U_t}{\mathrm{d}t} &= -\mathrm{i}\hat H_S(t)\hat U_t,
\end{align}
where $\hat H_S(t)$ is a Hermitian operator-valued function. Therefore, we have arrived at the expected exponential form of the evolution operator, generated by the \textit{Hamiltonian} of the system:
\begin{align}\label{eq:U_exp_form}
    \begin{array}{r}
    \textbf{Hilbert space representation}\\
    \text{(Dynamical law)}\\
    \end{array}\quad
    \hat U_{t,t'} &:= \hat U_t\hat U_{t'}^\dagger = \mathcal{T}\mathrm{e}^{-\mathrm{i}\int_{t'}^t\hat H_S(u)\mathrm{d}u}.
\end{align}

We conclude with a clarification: The method of ``initialization through measurement'' as the means of control over the initial conditions of the experiment implicitly assumes the \textit{stationarity} of probability distributions, i.e.,
\begin{align}\label{eq:stationarity_prob}
    P^{\tup{F}|\rho}_{\tup{t}|0}(\tup{f}) = P^{\tup{F}|\rho}_{t_n+t_0,\ldots,t_1+t_0|t_0}(\tup{f})\quad\text{for any $t_0>0$}.
\end{align} 
Otherwise, even with Markovianity at our disposal, it would not be possible to reset the ``state'' of the system, as the results of any given experiment would depend on the absolute time it started. In other words, if the results were to depend on the \textit{date} the experiment is taking place, then it would not be possible to compare the results of its past and future runs. But now that we have the detailed formal representation of the dynamics, we are obliged to determine how the requirement of stationarity translates into a constraint on the form of Hamiltonian.

When the stationarity property~\eqref{eq:stationarity_prob} is expressed within the Hilbert space representation,
\begin{align}
\nonumber
    \operatorname{tr}_S\Big[
            \Big(\prod_{j=n}^1\hat P^{F_j}_0(f_j)\hat U_{t_j+t_0,t_{j-1}+t_0}\Big)\hat\rho_0
            \Big(\prod_{j=1}^n\hat U_{t_j+t_0,t_{j-1}+t_0}^\dagger\hat P^{F_j}_0(f_j)\Big)
        \Big]&\\
    = \operatorname{tr}_S\Big[
            \Big(\prod_{j=n}^1\hat P^{F_j}_0(f_j)\hat U_{t_j,t_{j-1}}\Big)\hat\rho_0
            \Big(\prod_{j=1}^n\hat U_{t_j,t_{j-1}}^\dagger\hat P^{F_j}_0(f_j)\Big)
        \Big]&.
\end{align}
we conclude that the probabilities can be stationary only when the evolution operator satisfies the following condition:
\begin{align}
    \mathcal{T}\mathrm{e}^{-\mathrm{i}\int_{t_{j-1}+t_0}^{t_{j}+t_0}\hat H_S(u)\mathrm{d}u}
    =\mathcal{T}\mathrm{e}^{-\mathrm{i}\int_{t_{j-1}}^{t_{j}}\hat H_S(u-t_0)\mathrm{d}u}
    =\mathcal{T}\mathrm{e}^{-\mathrm{i}\int_{t_{j-1}}^{t_{j}}\hat H_S(u)\mathrm{d}u}\quad\text{for all $t_{j-1} < t_{j}$ and $t_0 >0$}.
\end{align}
It means that the Hamiltonian has to be time-independent, $\hat H_S(t) = \hat H_S$, in which case we obtain
\begin{align}
    \hat U_{t,t'} & = \sum_{n=0}^\infty (-\mathrm{i})^n\frac{(t-t')^n}{n!}\hat H_S^n
    = \mathrm{e}^{-\mathrm{i}(t-t')\hat H_S}.
\end{align}
Once we deduce how one might introduce the concept of external fields into the formalism, we shall revisit the physical viability of time-dependent Hamiltonians.

\section{Composite systems}\label{sec:composite_sys}

Previously, we deduced that every quantum system is formally characterized by a Hilbert space $\mathcal{H}_S$ and a Hamiltonian $\hat H_S$ that generates the evolution operator $\hat U_{t,t'}^S = {\exp}[-\mathrm{i}(t-t')\hat H_S]$ representing the system's dynamical law. We also established that bi-probabilities associated with observables of independent systems factorize in a manner analogous to probability distributions of independent stochastic variables, cf.~\ref{prop:bi-prob:factorization}. We can conclude from these facts that the Hilbert space representing a system composed of independent subsystems has the form of the tensor product of the Hilbert spaces of the constituents, 
\begin{equation}\mathcal{H}_{AB} = \mathcal{H}_A\otimes\mathcal{H}_B.\end{equation}
This means that the dynamical law of a system composed of independent subsystems is defined as the product of the evolution operators of its subsystems, 
\begin{equation}\hat U^{AB}_{t,t'} := \hat U^A_{t,t'}\otimes\hat U_{t,t'}^{B},\end{equation}
which implies 
\begin{equation}
    \hat H_{AB} = \hat H_A\otimes\hat 1_B + \hat 1_A\otimes\hat H_B.
\end{equation}
This tensor product structure extends to all operators involved in the Hilbert space representation: the metric representing the initialization event factorizes,  
\begin{equation}
    \hat\rho^{AB}_{t_0} =\hat\rho^A_{t_0}\otimes\hat\rho^B_{t_0},
\end{equation}
and the projectors associated with subsystem observables also have the product form. For instance, $\{\hat P^{F_A}(f)\otimes\hat 1_B\mid f\in\Omega(F_A)\}$ corresponds to the $A$-only observable measured by the $F_A$-device, $\{\hat 1_A\otimes\hat P^{F_B}(f)\mid f\in\Omega(F_B)\}$ corresponds to the $B$-only $F_B$-device, and thus, $\{\hat P^{F_AF_B}(f_a\wedge f_b) = \hat P^{F_A}(f_a)\otimes\hat P^{F_B}(f_b)\mid f_a\wedge f_b \in \Omega(F_AF_B) = \Omega(F_A)\times\Omega(F_B)\}$ corresponds to the observable measured by an $F_AF_B$-device, i.e., an $F_A$- an $F_B$-device deployed in tandem.

Therefore, two subsystems are mutually dependent when the initialization is not in a product form, $\hat\rho^{AB}_{t_0} \neq \hat\rho^A_{t_0}\otimes\hat\rho^B_{t_0}$, or when the Hamiltonian includes a \textit{coupling} operator, e.g., 
\begin{align}
    \hat H_{AB} = \hat H_A\otimes\hat 1_B + \hat 1_A\otimes\hat H_B + \lambda \hat V_A\otimes\hat V_B,
\end{align}
so that $\hat U^{AB}_{t,t'}\neq \hat U^A_{t,t'}\otimes\hat U^B_{t,t'}$. When the subsystems become dependent because of interactions, the influence of one system onto another can be understood by examining the bi-probabilities associated with observables of only one subsystem:
\begin{align}\label{eq:bi-prob_A}
    Q^{\tup{F}|\rho^A}_{\tup{t}|0}(\tup{f}^+,\tup{f}^-) &=
    \operatorname{tr}\Big[
            \Big(\prod_{j=n}^1\hat U^{AB}_{0,t_j}\hat P^{F_j}(f_j^+)\otimes\hat 1_B\hat U_{t_j,0}^{AB}\Big)\hat\rho^A_0\otimes\hat \rho^B_0
            \Big(\prod_{j=n}^1\hat U^{AB}_{0,t_j}\hat P^{F_j}(f_j^-)\otimes\hat 1_B\hat U_{t_j,0}^{AB}\Big)\Big],
\end{align}
where each $\{\hat P^{F_j}(f)\otimes\hat 1_B \mid f\in\Omega(F_j)\}$ corresponds to the observable of the subsystem $A$ measurable by the $F_j$-device.

To analyze the expression~\eqref{eq:bi-prob_A}, first let us take the complete set of orthogonal projectors in $B$
\begin{align}
    \left\{\hat P^B_t(b) = \mathrm{e}^{\mathrm{i}\hat H_B t}\hat P^B_0(b)
        \mathrm{e}^{-\mathrm{i}\hat H_B t}
        \mid b\in\Omega(B)\right\}    
\end{align}
obtained as the solution to the eigenproblem of the $B$-side coupling operator $\hat V_B$:
\begin{align}\label{eq:VB_eigenproblem}
    \hat P^B_t,v_B:\quad \big(\hat V_B(t) - v_B(b)\hat 1_B\big)\hat P^B_t(b) = 0;
\end{align}
where $\hat V_{A/B}(t) = \exp(\mathrm{i}\hat H_{A/B}t)\hat V_{A/B}\exp(-\mathrm{i}\hat H_{A/B}t)$. In principle, these projectors correspond to the $B$-device that could be deployed to measure an observable belonging to subsystem $B$. Thus, as per property~\ref{prop:bi-prob:master_measure}, we can associate with this observable the bi-trajectory measure $\mathcal{Q}^{B|\rho^B}$ such that
\begin{align}
    \iint \Big(\prod_{j=1}^n\delta_{b^+(t_j),b_j^+}\delta_{b^-(t_j),b_j^-}\Big)\bimeasure[B|\rho^B]{b}
    =\operatorname{tr}_B\Big[\Big(\prod_{j=n}^1\hat P^{B}_{t_j}(b_j^+)\Big)\hat\rho^B_0\Big(\prod_{j=1}^n\hat P_{t_j}^{B}(b_j^-)\Big)\Big],
\end{align}
Next, take note of the following identity: given a sequence of timings $\bm{s}_n = (s_n,\ldots,s_1)$ split into two blocks $I^+$ and $I^-$ such that $\{n,\ldots,1\} = I^+\cup I^-$, the bi-probabilities satisfy
\begin{align}
\nonumber
    &\operatorname{tr}_B\Big[
    \mathcal{T}\Big\{\prod_{j\in I^+}\hat V_B(s_j)\Big\}\hat\rho^B_0
    \Bigl(\mathcal{T}\Big\{\prod_{k\in I^-}\hat V_B(s_k)\Big\}\Bigr)^\dagger
    \Big]\\
\nonumber
    &\phantom{=}=
    \operatorname{tr}_B\Big[
    \mathcal{T}\Big\{\prod_{j\in I^+}\hat V_B(s_j)\hat 1_B\Big\}\hat\rho^B_0
    \Bigl(\mathcal{T}\Big\{\prod_{k\in I^-}\hat V_B(s_k)\hat 1_B\Big\}\Bigr)^\dagger
    \Big]\\
\nonumber
    &\phantom{=}=\operatorname{tr}_B\Big[
    \mathcal{T}\Big\{\prod_{j\in I^+}\hat V_B(s_j)\sum_{b^+_j}\hat P_{s_j}^B(b_j^+)\Big\}\hat\rho^B_0
    \Bigl(\mathcal{T}\Big\{\prod_{k\in I^-}\hat V_B(s_k)\sum_{b_k^-}\hat P^B_{s_k}(b_k^-)\Big\}\Bigr)^\dagger
    \Big]\\
\nonumber
    &\phantom{=}=\operatorname{tr}_B\Big[
    \mathcal{T}\Big\{
        \prod_{j\in I^+}\sum_{b_j^+}v_B(b_j^+)\hat P^B_{s_j}(b_j^+)\Big\}
    \hat\rho^B_0
    \Bigl(\mathcal{T}\Big\{
        \prod_{k\in I^-}\sum_{b_k^-}v_B(b_k^-)\hat P^B_{s_k}(b_k^-)\Big\}\Bigr)^\dagger
    \Big]\\
\nonumber
    &\phantom{=}=\operatorname{tr}_B\Big[
    \Big(\prod_{i=n}^1\sum_{b_i^+}
        \hat P^B_{s_i}(b_i^+)\Big\{\prod_{j\in I^+}v_B(b_j^+)\Big\}\Big)
    \hat\rho^B_0
    \Big(\prod_{i=1}^n\sum_{b_i^-}\hat P^B_{s_i}(b_i^-)\Big\{\prod_{k\in I^-}v_B(b_k^-)\Big\}\Big)
    \Big]\\
\nonumber
    &\phantom{=}=\sum_{\bm{b}^\pm_n}\Big(
        \prod_{j\in I^+}v_B(b_j^+)\prod_{k\in I^-}v_B(b^-_k)
    \Big)\operatorname{tr}_B\Big[
        \Big(\prod_{i=n}^1\hat P^B_{s_i}(b_i^+)\Big)\hat\rho_0^B
        \Big(\prod_{i=1}^n\hat P^B_{s_i}(b_i^-)\Big)
    \Big]\\
\label{eq:moments}
    &\phantom{=}=\iint\Big(
        \prod_{j\in I^+}v_B\big(b^+(s_j)\big)
        \prod_{k\in I^-}v_B\big(b^-(s_k)\big)
    \Big)\bimeasure[B|\rho^B]{b}.
\end{align}
We can now leverage this relation to express the bi-probability associated with $A$-observables as an average over the bi-trajectory $(\,b^+\!\!,b^-)$~\cite{Szankowski_SciPostLecNotes23,Szankowski_Quantum24,szankowski-2025}:
\begin{align}
\nonumber
    Q_{\tup{t}|0}^{\tup{F}|\rho^A}(\tup{f}^+,\tup{f}^-) &=
    \operatorname{tr}\Big[\Big(\prod_{j=n}^1
        \mathrm e^{\mathrm i \hat H_{AB}t_j}
        \hat P^{F_j}(f_j^+)\otimes\hat 1_B
        \mathrm e^{-\mathrm i \hat H_{AB}t_j}
        \Big)
        \hat\rho_0^A\otimes\hat\rho_0^B\\
\nonumber
    &\phantom{=\operatorname{tr}\Big[}\times
        \Big(\prod_{j=1}^n
        \mathrm e^{\mathrm i \hat H_{AB}t_j}
        \hat P^{F_j}(f_j^-)\otimes\hat 1_B
        \mathrm e^{-\mathrm i \hat H_{AB}t_j}
        \Big)\Big]\\
\nonumber
    &=\operatorname{tr}\Big[
        \Big(
            \prod_{j=n}^1\mathrm{e}^{\mathrm{i}\hat H_A t_j}
            \hat P^{F_j}(f_j^+)\mathrm{e}^{-\mathrm{i}\hat H_A t_j}\otimes\hat 1_B
            \,\mathcal{T}\mathrm{e}^{
            -\mathrm{i}\lambda\int_{t_{j-1}}^{t_j}
            \hat V_A(s)\otimes\hat V_B(s)\mathrm{d}s
            }
        \Big)
        \hat\rho^A_0\otimes\hat\rho^B_0\\
\nonumber
    &\phantom{=\operatorname{tr}\Big[}\times
    \Big(\prod_{j=1}^n \Big(\mathcal{T}\mathrm{e}^{
        -\mathrm{i}\lambda\int_{t_{j-1}}^{t_j}\hat V_A(s)\otimes\hat V_B(s)
        \mathrm{d}s
    }\Big)^\dagger\mathrm{e}^{\mathrm{i}\hat H_A t_j}\hat P^{F_j}(f_j^-)
    \mathrm{e}^{-\mathrm{i}\hat H_A t_j}\otimes\hat 1_B\Big)
    \Big]\\
\nonumber
    &=\iint\operatorname{tr}\Big[
        \Big(\prod_{j=n}^1\hat P^{F_j}_{t_j}(f_j^+)
            \,\mathcal{T}\mathrm{e}^{
                -\mathrm{i}\lambda\int_{t_{j-1}}^{t_j}
                v_B(b^+(s))\hat V_A(s)\mathrm{d}s
            }
        \Big)\hat\rho^A_0\\
\nonumber
    &\phantom{=\operatorname{tr}\Big[}\times
    \Big(\prod_{j=1}^n \Big(\mathcal{T}\mathrm{e}^{
        -\mathrm{i}\lambda\int_{t_{j-1}}^{t_j}v_B(b^-(s))\hat V_A(s)
        \mathrm{d}s
    }\Big)^\dagger\hat P^{F_j}_{t_j}(f_j^-)\Big)
    \Big]\bimeasure[B|\rho^B]{b}\\
\label{eq:open_bi-prob}
    &= \iint \operatorname{tr}_A\Big[
        \Big(\prod_{j=n}^1\hat P^{F_j}_{t_j}[b^+](f_j^+)\Big)\hat\rho^A_0
        \Big(\prod_{j=1}^n\hat P^{F_j}_{t_j}[b^-](f_j^-)\Big)
    \Big]
    \bimeasure[B|\rho^B]{b},
\end{align}
where the bi-trajectory-dependent projectors are defined as:
\begin{align}\label{eq:external_field}
    \hat P^{F_j}_{t}[h](f) &:= 
    \mathcal{T}\exp\Big[{-\mathrm{i}}
    \int_0^t(\hat H_A + \lambda v_B( h(s) )\hat V_A)\mathrm{d}s\Big]^\dagger
    \hat P^{F_j}(f)
    \mathcal{T}\exp\Big[{-\mathrm{i}}\int_0^t(\hat H_A + \lambda v_B(h(s))\hat V_A)\mathrm{d}s\Big],
\end{align}
and the identity~\eqref{eq:moments} was used on each term of the time-ordered exponential series expansion to `replace' the $B$-side operators with the corresponding component of bi-trajectory.

We take note of two important features of Eq.~\eqref{eq:open_bi-prob}. The first is that the unitary evolution operators in Eq.~\eqref{eq:external_field} are generated by the time-dependent Hamiltonian $\hat H[h](s) = \hat H_A + \lambda v_B(h(s))\hat V_A$. The real-valued function $v_B(h(s))$, which introduces a time-dependence into the Hamiltonian, can be interpreted as an external field driving the dynamics via coupling with the operator $\hat V_A$. Of course, in Eq.~\eqref{eq:open_bi-prob}, the role of the external driver is played by the components of the bi-trajectory $(b^+(s),b^-(s))$: $b^+(s)$ is coupled to the sequence progressing forwards in time, while $b^-(s)$ drives the sequence going backwards in time, thereby facilitating the quantum interference between these two branches of the evolution. 

In the special case where $\bimeasure[B|\rho^B]{b} = \delta[b^+-b^0]\delta[b^--b^0][\mathcal{D}b^+][\mathcal{D}b^-]$, meaning that both components of the bi-trajectory overlap with the one trajectory $t\mapsto b^0(t)$, the influence from $B$ takes on the form identical to the driving with the given external classical force field~\cite{Szankowski_SciRep20}. Therefore, the system $A$, that is in fact open to interactions with system $B$, can be formally described in this scenario as a closed system with the dynamical law generated by the time-dependent Hamiltonian $\hat H_A(s) = \hat H_A + \lambda v_B(b^0(s))\hat V_A$.

The second important feature is that the projectors $\{\hat P_t^B(b)\mid b\in\Omega(B)\}$ corresponding to a $B$-only observable, which we associate with the bi-trajectory measure $\mathcal{Q}^{B|\rho^B}$, originate from the formal spectral decomposition of the coupling operator $\hat V_B$ (cf. Eq.~\eqref{eq:moments}). Specifically, given the solution of the eigenproblem~\eqref{eq:VB_eigenproblem} we have
\begin{align}
    \hat V_B(t) =
    \hat V_B(t)\hat 1_B = \hat V_B(t)\sum_{b\in\Omega(B)}\hat P^B_t(b) =
    \sum_{b\in\Omega(B)}v_B(b) \hat P_t^{B}(b) \equiv \sum_{v\in\Omega(V_B)}v \hat P^{V_B}_t(v).
\end{align}
Generalizing this observation, any Hermitian operator $\hat F=\sum_{f\in\Omega(F)}f\hat P^F(f)$ can be understood as the Hilbert space representation of the observable probed by a measuring device associated with the set of projectors $\{\hat P^F(f) \mid f\in\Omega(F)\}$. Then, it follows that quantum systems influence each other by means of the dynamics generated by the coupling of their observables---the observables that could be in principle probed by the measuring device deployed by the classical observer. Thus, we establish the familiar notion, known from the standard formalism, that the observables of quantum systems correspond to Hermitian operators acting in the system Hilbert space.

\section{The master object of the formalism}\label{sec:ture_master}

When reviewing the structure of the standard formalism of quantum mechanics in Part I, we observed that the quantum state represented by a density matrix $\hat\rho(t)$ can be considered as a \textit{master object} of this formalism: For as long as one restricts their interests to single-measurement experiments (ignoring, for the moment, the issue of initialization), then, indeed, $\hat\rho(t)$ is all that is needed in order to fully describe all the observable facts. As it was noted in Section~\ref{sec:obs--projector_link}, even the passage of time can be reduced in this setting to the observer choosing an appropriate measuring device. Hence, the single-measurement context can be considered as quantum statics, analogous in its scope to, e.g., hydrostatics of classical fluid mechanics. It is then not surprising that for quantum statics, similarly to its classical analogues, it is a natural choice to designate the system state as the master object of the formalism. 

Moving beyond ``quantum statics'' to ``quantum dynamics'' which involve scenarios such as sequential measurements, interacting quantum systems (cf.~Section~\ref{sec:composite_sys}), etc., the state $\hat\rho(t)$ is dethroned by bi-probabilities as the master object, and the role of the state, corresponding here to a metric matrix, is reduced to a convenient representation of the initialization event. In the setting where only a single observable $F$ is relevant, as indicated by the property~\ref{prop:bi-prob:master_measure}, all bi-probabilities $Q^{F|\rho}_{\bm t_n|t_0}$ associated with $F$ are simply the discrete-time restrictions of the bi-trajectory measure $\bimeasure[F|\rho]{f}$. It is thus revealed that, in this limited---but still dynamical---context, the status of master object belongs to the bi-trajectory measure, as it is the source of all other objects in the formalism. 

However, unlike classical theories, quantum systems accommodate an infinite number of inequivalent observables, and in the general context, the formalism must utilize bi-probabilities $Q_{\tup{t}|t_0}^{\tup{F}|\rho}$ associated with multiple observables. Therefore, just like the state before it, the bi-trajectory measure $\bimeasure[F|\rho]{f}$ surely has to be dethroned by something else. A natural candidate would be a bi-trajectory measure that extends \textit{all} multi-observable bi-probabilities; the question is if such an object actually exists.

To answer this question, we must first introduce a higher level of abstraction to the formalism by defining \textit{system} bi-probabilities that are not associated with any observable or initialization event. To begin with, let us choose an arbitrary \textit{reference} basis in the system Hilbert space $\mathcal{H}_S$,
\begin{align}
    \mathbb{E} = \left\{ |\eta\rangle\in\mathcal{H}_S \mid \eta \in \{1,\ldots,d\}\,,\, \langle \eta|\eta'\rangle = \delta_{\eta,\eta'}\right\},
\end{align}
where $d=\operatorname{dim}\mathcal{H}_S$. Note that, for any fine-grained observable $K$ associated with the corresponding set of rank-$1$ orthogonal projectors
\begin{align}
    \hat P^K_0(k)=|\Psi^K_0(k)\rangle\langle\Psi^K_0(k)|\ :\ 
    \mathbb{K}_0 = \big\{|\Psi^K_0(k)\rangle\in\mathcal{H}_S\mid k\in\Omega(K), \langle\Psi_0^K(k)|\Psi^K_0(k')\rangle=\delta_{k,k'}\big\},
\end{align}
there always exists an unitary basis transformation $\mathbb{K}_0\leftrightarrow\mathbb{E}$:
\begin{align}
    \mathrm{e}^{\mathrm{i}\vec{\hat T}\cdot\vec{\varphi}_K}
    |\eta\rangle
    = \exp\left(\mathrm{i}\sum_{\ell=0}^{d^2-1} \hat T_\ell\varphi^\ell_K\right)
    |\eta\rangle
    = |\Psi^K_0\big(k(\eta)\big)\rangle.
\end{align}
Here, we chose to parameterize the unitary operators in $\mathcal{H}_S$ using the set of Hermitian traceless Lie algebra generators $\vec{\hat T} = (\hat T_1,\ldots,\hat T_{d^2-1})$ and $\hat T_0 = \hat 1$, and coordinates on a $(d^2-1)$-dimensional manifold $\mathbb{S}^{d^2-1}$, $\vec\varphi = (\varphi^1,\ldots,\varphi^{d^2-1})$. Such a transformation is unique up to permutations of basis elements, hence the need for the map $k: \{1,\ldots,d\} \to \Omega(K)$ that selects a concrete correspondence between the reference basis indexes and the observable's spectrum $\Omega(K)$.

In the general case of an observable $F$ associated with projectors $\{\hat P^F_0(f')\mid f'\in\Omega(F)\}$, that is not necessarily perfectly fine-grained, the transformation coordinates and the index mapping can be found by solving the corresponding eigenproblem:
\begin{align}
    \vec\varphi_F\in \mathbb S^{d^2-1},\,
    %f:\{1,\ldots,d\}\to\Omega(F)
    \eta\mapsto f(\eta)\in\Omega(K)
    \quad:\quad
    \left(\sum_{f'\in\Omega(F)}f'\hat P^F_0(f') - f(\eta)\hat 1\right)
    \mathrm{e}^{\mathrm{i}\vec{\hat T}\cdot\vec{\varphi}_F}|\eta\rangle = 0.
\end{align}

Composing the basis transformation with the temporal evolution operator generated by the system Hamiltonian $\hat H_S$ allow us to parameterize any observable-associated projector with the `space-time' coordinates $\tau =(t,\vec\varphi)$:
\begin{align}
    \hat P^{F}_t(f') 
    &= \sum_{\eta=1}^d \delta_{f',f(\eta)}
        \mathrm{e}^{\mathrm{i}\hat H_S t}
        \mathrm{e}^{\mathrm{i}\vec{\hat T}\cdot\vec{\varphi}_F}
        |\eta\rangle\langle\eta|
        \mathrm{e}^{-\mathrm{i}\vec{\hat T}\cdot\vec{\varphi}_F}
        \mathrm{e}^{-\mathrm{i}\hat H_S t}
    \equiv
        \sum_{\eta=1}^d \delta_{f',f(\eta)}\hat P_{(t,\vec\varphi_F)}(\eta).
\end{align}
This parameterization can now be used to define a family of bi-probabilities which are labeled with a sequence of `space-time' coordinates $\tau_j = (t_j,\vec{\varphi}_j)$, rather than observables, time, and the initial condition:
\begin{align}\label{eq:joint_sys_bi-prob}
    Q_{\tup{\tau}|\tau_0}(\tup{\eta}^+,\tup{\eta}^-|\eta_0^+,\eta_0^-) := \operatorname{tr}\Big[
        \Big(\prod_{j=n}^0 \hat P_{\tau_j}(\eta^+_j)\Big)
        \Big(\prod_{j=0}^n \hat P_{\tau_j}(\eta^-_j)\Big)
    \Big].
\end{align}
These system bi-probabilities $Q_{\tup{\tau}|\tau_0}$ are the master objects with respect to their observable-associated cousins $Q_{\tup{t}|t_0}^{\tup{F}|\rho}$, as the latter decompose into convex combinations of the former,
\begin{align}\label{eq:sys-bi-prob:observables}
    Q_{\tup{t}|t_0}^{\tup{F}|\rho}(\tup{f}^+,\tup{f}^-) &=
        \sum_{\tup{\eta}^\pm,\eta_0}\Big(\prod_{j=1}^{n}
            \delta_{f_{j}(\eta_{j}^+),f_{j}^+}\delta_{f_{j}(\eta_{j}^-),f_{j}^-}\Big)
        Q_{(t_n,\vec\varphi_n),\ldots,(t_1,\vec\varphi_1)|(t_0,\vec\varphi_0)}(\tup{\eta}^+,\tup{\eta}^-|\eta_0,\eta_0)\rho(\eta_0),
\end{align}
where the `spatial'-coordinates $\vec\varphi_n,\ldots,\vec\varphi_1$ and $\vec\varphi_0$ and the index-to-spectrum mappings $f_j,\rho$ are each found by solving the corresponding eigenproblem.

It is easy to verify that the functions $Q_{\tup{\tau}|\tau_0}$ satisfy all the properties~\ref{prop:bi-prob:norm}--\ref{prop:bi-prob:bi-consistency}, including the crucial bi-consistency. Therefore, the natural next step would be to extend the family of system bi-probabilities to a bi-trajectory measure, as it was done in Part I for single-observable $Q_{\tup{t}|t_0}^{F|\rho}$ (see property~\ref{prop:bi-prob:master_measure}); that is, we suppose that
\begin{align}\label{eq:sys-bi-prob:extension}
    Q_{\tup{\tau}|\tau_0}(\tup{\eta}^+,\tup{\eta}^-|\eta_0^+,\eta_0^-)
    &= \iint \Big(\prod_{j=0}^n\delta_{\eta^+_j,\eta^+(\tau_j)}\delta_{\eta^-_j,\eta^-(\tau_j)}\Big)\bimeasure{\eta},
\end{align}
where $\bimeasure{\eta}$ is a complex-valued measure on the space of bi-trajectories,
\begin{align}\label{eq:bi-field}
    \mathbb{R}\times \mathbb S^{d^2-1}\ni(t,\vec\varphi)\mapsto(\eta^+(t,\vec\varphi),\eta^-(t,\vec\varphi))\in\{1,\ldots,d\}\times\{1,\ldots,d\}.
\end{align}
Then, assuming such a $\mathcal{Q}$ does exist, it would relate to the single-observable master measure $\mathcal{Q}^{F|\rho}$ from~\ref{prop:bi-prob:master_measure} in the following way:
\begin{align}
\nonumber
    Q^{F|\rho}_{\tup{t}|0}(\tup{f}^+,\tup{f}^-)
    &=\iint \Big(\prod_{j=1}^n 
        \delta_{f_j^+,f^+(t_j)}\delta_{f_j^-,f^-(t_j)}
    \Big)\bimeasure[F|\rho]{f}\\
\nonumber
    &=\iint\Big(\prod_{j=1}^n
        \delta_{f_j^+,f(\eta^+(t_j,\vec\varphi_F))}
        \delta_{f_j^-,f(\eta^-(t_j,\vec\varphi_F))}
    \Big)\\
\label{eq:sys_vs_obs}
    &\phantom{=\iint}\times
    %\sqrt{\rho(\eta^+(0,\vec\varphi_0))\rho(\eta^-(0,\vec\varphi_0))}\,
    \Big(\sum_{\eta_0=1}^d \rho(\eta_0)
        \delta_{\eta_0,\eta^+(0,\vec\varphi_0)}
        \delta_{\eta_0,\eta^-(0,\vec\varphi_0)}
    \Big)
    \bimeasure{\eta},
\end{align}
where $\hat F=\sum_\eta f(\eta)\hat P_{(0,\vec\varphi_F)}(\eta)$ and $\hat\rho_0 = \sum_\eta \rho(\eta)\hat P_{(0,\vec\varphi_0)}(\eta)$; therefore, $\mathcal{Q}^{F|\rho}$ can be formally considered a \textit{pushforward} of measure $\mathcal{Q}$.

Unfortunately, for several non-trivial reasons (see~\cite{Lonigro_Quantum24} for a detailed explanation), we cannot prove the existence of the master bi-trajectory measure $\mathcal{Q}$ with the same method used to prove ~\ref{prop:bi-prob:master_measure}. Nevertheless, even in the absence of a formal proof, we can still formulate a strong supporting argument:

Assuming that the master measure exists, we can define the following super-operator:
\begin{align}
\nonumber
    \Lambda_t &:= \iint \Big(\mathcal{T}\mathrm{e}^{-\mathrm{i}\int_0^t\hat H[\eta^+](s)\mathrm{d}s}\Big)\bullet\Big(\mathcal{T}\mathrm{e}^{-\mathrm{i}\int_0^t\hat H[\eta^-](s)\mathrm{d}s}\Big)^\dagger\\
\label{eq:bi-average_dynamical_map}
    &\phantom{:=\iint}\times
    \Big(\sum_{\eta=1}^d\rho(\eta_0)
        \delta_{\eta_0,\eta^+(0,\vec\varphi_0)}
        \delta_{\eta_0,\eta^-(0,\vec\varphi_0)}\Big)
    \bimeasure{\eta}
\end{align}
with an arbitrary probability distribution $\rho(\eta_0)$ and
\begin{align}
    \hat H[\eta](s) &:= \hat H_0 + \sum_{\alpha=1}^N f_\alpha\big(\eta(s,\vec\varphi_\alpha)\big)\hat H_\alpha,
\end{align}
where $\{\hat H_\alpha \mid \alpha = 0,1,\ldots,N\}$ is a set of arbitrary Hermitian operators in some finite-dimensional Hilbert space $\mathcal{H}_O$, and
\begin{align}
    \Big\{\hat F_\alpha = \sum_{\eta=1}^d f_\alpha(\eta)\hat P_{(0,\vec\varphi_\alpha)}(\eta) \ \Big|\ \alpha = 1,\ldots,N\Big\}    
\end{align}
is an arbitrary set of observables in $\mathcal{H}_S$.

Such $\Lambda_t$ is a \textit{characteristic function} of the measure $\bimeasure{\eta}$ which generates moments in the form of bi-probabilities~\cite{Szankowski_SciPostLecNotes23}:
\begin{align}
\nonumber
    \Lambda_t  &=\big(
        \mathrm{e}^{-\mathrm{i}\hat H_0 t}\bullet\mathrm{e}^{\mathrm{i}\hat H_0 t}
    \big)
    \sum_{n=0}^\infty(-\mathrm{i})^n\int_0^t\mathrm{d}s_n\cdots\int_0^{s_2}\mathrm{d}s_1\sum_{\alpha_n,\ldots,\alpha_1}\\
\nonumber
    &\phantom{==}\times
        \sum_{\bm{\eta}_n^\pm}\sum_{\eta_0}Q_{(s_n,\vec\varphi_{\alpha_n}),\ldots,(s_1,\vec\varphi_{\alpha_1})|(0,\vec\varphi_0)}(\tup{\eta}^+,\tup{\eta}^-|\eta_0,\eta_0)\rho(\eta_0)\\
\nonumber
    &\phantom{===}\times
        \prod_{j=n}^1\big(
            f_{\alpha_j}(\eta_j^+)
            \mathrm{e}^{\mathrm{i}\hat H_0 s_j}\hat H_{\alpha_j}
            \mathrm{e}^{-\mathrm{i}\hat H_0 s_j}\bullet
            -\bullet\mathrm{e}^{\mathrm{i}\hat H_0 s_j}
            \hat H_{\alpha_j}\mathrm{e}^{-\mathrm{i}\hat H_0 s_j}
            f_{\alpha_j}(\eta_j^-)\big).
\end{align}

On the other hand, using the identity analogous to Eq.~\eqref{eq:moments} one shows that $\Lambda_t$ is also a completely positive and trace-preserving (CPTP) dynamical map well known from the standard theory of open quantum systems,
\begin{align}
    \Lambda_t\hat\rho^O_0 = 
    \operatorname{tr}_S\Big[\mathrm{e}^{-\mathrm{i}t\hat H_{OS}}
    \hat\rho^O_0\otimes\Big(\sum_{\eta=1}^d \rho(\eta)\hat P_{(0,\vec\varphi_0)}(\eta)\Big)
    \,\mathrm{e}^{\mathrm{i}t\hat H_{OS}}\Big],
\end{align}
where the system $S$ play the role of the environment of open system $O$, and the total $OS$ Hamiltonian was set to
\begin{align}
    \hat H_{OS} = \hat H_0\otimes\hat 1_S + \hat 1_O\otimes\hat H_S + \sum_{\alpha=1}^N\hat H_\alpha \otimes \Big(\sum_{\eta=1}^d f_\alpha(\eta)\hat P_{(0,\vec\varphi_\alpha)}(\eta)\Big).
\end{align}

Therefore, the characteristic function of $\bimeasure{\eta}$ is a well-behaved, regular CPTP map, and thus, the underlying bi-trajectory measure should also be well-behaved. This concludes the argument.

The system bi-trajectory measure $\bimeasure{\eta}$ is the true master object, as the relations~\eqref{eq:sys-bi-prob:extension} and~\eqref{eq:sys-bi-prob:observables} show that it is the source of all the elements of the formalism. The master measure is not associated with any particular observable or initial condition, but rather it is associated with the quantum system itself. Indeed, the master measure is solely determined by the Hilbert space $\mathcal{H}_S$ of the system and its Hamiltonian $\hat H_S$, and, as previously discussed, the system is also formally identified by the same pair. Thus, in the bi-trajectory formalism, the system is synonymous with its bi-trajectory measure $\bimeasure{\eta}$.

\section{Conclusions}\label{sec:conclusions}

We have demonstrated how the bi-trajectory formalism for quantum theory is deduced from the phenomenology of experiments involving a sequential deployment of measuring devices. In the bi-trajectory framework, a quantum system is identified with a $d$-dimensional Hilbert space $\mathcal{H}$ and a hermitian operator $\hat H$---the system's Hamiltonian. The dynamical properties of the system are encapsulated by a complex-valued measure on the space of bi-trajectories:
\begin{align*}
    \text{\bf Quantum System}:&\quad\bimeasure[(\mathcal{H},\hat H)]{\eta};\\
    \text{\textbf{Bi-trajectory}}:&\quad
    (t,\vec\varphi)
    \mapsto\big(\,\eta^+(t,\vec\varphi)\,,\,\eta^-(t,\vec\varphi)\,\big),
\end{align*}
where $t$ represents time, and the coordinates $\vec\varphi$ parameterize transformations of the reference basis. The bi-trajectory measure itself is \textit{generated} by the Hamiltonian, in the sense that the value of any discrete restriction,
\begin{align}
\nonumber
    &\iint \Big(\prod_{j=0}^n
        \delta_{\eta^+_j,\eta^+(t_j,\vec\varphi_j)}
        \delta_{\eta^-_j,\eta^-(t_j,\vec\varphi_j)}
    \Big)\bimeasure[(\mathcal{H},\hat H)]{\eta}\\
    &\phantom{=}=\operatorname{tr}\Big[
        \Big(\prod_{j=n}^0
        \mathrm{e}^{\mathrm{i}\hat H t_j}
        \mathrm{e}^{\mathrm{i}\vec{\hat T}\cdot\vec\varphi_j}|\eta_j^+\rangle
        \langle\eta_j^+|\mathrm{e}^{-\mathrm{i}\vec{\hat T}\cdot\vec\varphi_j}
        \mathrm{e}^{-\mathrm{i}\hat H t_j}
        \Big)
        \Big(\prod_{j=0}^n
        \mathrm{e}^{\mathrm{i}\hat H t_j}
        \mathrm{e}^{\mathrm{i}\vec{\hat T}\cdot\vec\varphi_j}|\eta_j^-\rangle
        \langle\eta_j^-|\mathrm{e}^{-\mathrm{i}\vec{\hat T}\cdot\vec\varphi_j}
        \mathrm{e}^{-\mathrm{i}\hat H t_j}
        \Big)
    \Big]
\end{align}
is fully determined by $\hat H$ defining the system's dynamical law (in contrast, the generators of basis transformations are fixed by the arbitrary choice of reference basis)

By design, the deduced formal description is necessarily consistent with the empirical observations it was derived from. However, the interpretation of the elements of this formalism does not have to (and usually does not) align with the interpretations of other formulations, including the standard formalism. This situation is analogous to classical mechanics in its various forms. Over the years, the formalism of the classical theory has been restructured in numerous ways: starting from the formulation based on Newton's Laws of Motion, and progressing into more elegant phase-space-based formalisms, exemplified most famously by Hamilton's formulation. Ostensibly, all these formulations describe the same \textit{physical theory} and are in agreement regarding their empirically testable predictions (within the domain of classical physics). Yet, the formulations are structured very differently, leading to different interpretations and, consequently, different \textit{explanations}. For instance, Hamilton's reformulation was motivated by the quest to eliminate the problematic concept of \textit{force} from the theory~\cite{Poincare_05}. Hence, while Newton's formulation explains all motion through the application and balancing of forces, Hamilton's formalism explains the same motion without reference to such concepts. Consequently, the two formulations of the theory are \textit{not} equivalent, as none of their constituent elements (the formalism, its interpretation, and the resultant explanations) are in exact one-to-one correspondence. However, this does not negate the fact they agree in their testable predictions; rather, it should be taken to mean that neither formulation is made redundant by the other---each has the potential to improve in its own unique way our understanding of the theory they describe.

The precedence set by classical mechanics shows that a specific formalism of a physical theory need not be its defining feature. In other words, a theory can retain its identity even when its formal description is replaced by a vastly different one. We believe that the bi-trajectory formalism stands in relation to the standard formalism of quantum mechanics much as Hamilton's formulation does to Newtonian mechanics, rather than as Einstein's theory of special relativity does to classical mechanics. In other words, we view the bi-trajectory formalism as a reformulation of quantum mechanics, not as the formalism of a new theory that would generalize, or be generalized by, the standard quantum theory. 

Nevertheless, the bi-trajectory formalism is not equivalent to the standard formalism; the two formulations are structured very differently, and thus, must be interpreted differently as well. In the standard formulation, with its dichotomous mode of description, the quantum state,
\begin{align}
    \text{\textbf{Quantum state} (standard formalism)}:\quad
    \hat \rho(t) = \mathrm{e}^{-\mathrm{i}\hat Ht}\hat\rho_{0}\mathrm{e}^{\mathrm{i}\hat Ht},
\end{align}
is considered in most interpretations as the master object~\cite{Letertre_21}. In the bi-trajectory formalism, the bi-trajectory measure $\mathcal{Q}^{(\mathcal{H},\hat H)}$ is the master object, and the state has no formal analogue. Instead, we recognize that, formally, the operator
\begin{align*}
    \hat\rho(0) &= \hat\rho_0 = \sum_{\eta=1}^d \rho(\eta)
        \mathrm{e}^{\mathrm{i}\vec{\hat T}\cdot\vec\varphi_0}|\eta\rangle
        \langle \eta|\mathrm{e}^{-\mathrm{i}\vec{\hat T}\cdot\vec\varphi_0},
\end{align*}
could be utilized in the appropriate context as a metric matrix representing some initialization event. However, even in realist interpretations, the quantum state of the standard formalism, in and of itself, does not correspond to any empirically testable prediction. The standard formalism constructs testable predictions using the state as a component of a larger formal expression. For example, probabilities describing results of measurements are obtained with the Born rule. In general, predictions in the standard formulation (including the Born rule) take the form of functions of \textit{multi-time correlations}~\cite{szankowski-2025}:
\begin{align}
    \text{\textbf{Multi-time correlation} (std. formalism)}:\ 
    \operatorname{tr}\Big[
        \mathcal{T}\big\{\prod_{j\in I^+}\hat F_j(t_j)\big\}
        \,\hat\rho(0)\,
        \Bigl(\mathcal{T}\big\{\prod_{k\in I^-}\hat F_k(t_k)\big\}\Bigr)^\dagger
    \Big],
\end{align}
where $t_n > \cdots > t_1 > 0$, $I^+\cup I^- = \{ n,\ldots, 1\}$, and the hermitian operators
\begin{align}
    \hat F_j(t_j) &= \sum_{\eta=1}^d f_j(\eta)
        \mathrm{e}^{\mathrm{i}\hat Ht_j}
        \mathrm{e}^{\mathrm{i}\vec{\hat T}\cdot\vec\varphi_j}|\eta\rangle
        \langle\eta|\mathrm{e}^{-\mathrm{i}\vec{\hat T}\cdot\vec\varphi_j}
        \mathrm{e}^{-\mathrm{i}\hat H t}
\end{align}
represent the Heisenberg picture of observables being correlated. According to the identity~\eqref{eq:moments}, these correlation functions correspond to moments of bi-trajectory measure:
\begin{align}
\nonumber
    \operatorname{tr}\Big[
        \mathcal{T}\big\{\prod_{j\in I^+}\hat F_j(t_j)\big\}
        \hat\rho(0)
        \Bigl(\mathcal{T}\big\{\prod_{k\in I^-}\hat F_k(t_k)\big\}\Bigr)^\dagger
    \Big] &= \iint 
        \Big(\prod_{j\in I^+}f_j\big(\eta^+(t_j,\vec\varphi_j)\big)
        \prod_{k\in I^-}f_k\big(\eta^-(t_k,\vec\varphi_k)\big)\Big)\\
\nonumber
    &\phantom{=\iint}\times
        \Big(\sum_{\eta_0}\rho(\eta_0)
            \delta_{\eta_0,\eta^+(0,\vec\varphi_0)}
            \delta_{\eta_0,\eta^-(0,\vec\varphi_0)}
        \Big)\\
\label{eq:mtc}
    &\phantom{=\iint\times}\times
        \bimeasure[(\mathcal{H},\hat H)]{\eta},
\end{align}
and thus, the standard and the bi-trajectory formulation are in perfect alignment in regards to the empirically testable predictions available to the standard formalism.

A comprehensive restructuring of the formalism offers a relatively rare opportunity to reassess the physical and philosophical meaning of the theory from a fresh perspective. As the example of classical mechanics illustrates, though different formulations are often not equivalent (in terms of structure, interpretation, and explanations), it can be advantageous to switch between them depending on the context; it is possible that some fundamental problems with the theory that seem insurmountable within one formalism might have a solution---or cease to be problems---within a different formal framework. One such problematic aspect of the standard quantum mechanics could be the \textit{measurement problem}. Most---if not all---conceptual and practical issues surrounding measurement in quantum mechanics remain unresolved because the standard formulation, with its dichotomous structure, makes modeling the process of measurement impossible~\cite{Caves_PRD1986}. Indeed, for a model of measuring device to be successful it needs a formal description that does not break when transitioning between measurement and non-measurement contexts---due to the Heisenberg cut, such models are simply outside the scope of the standard formalism~\cite{Frauchiger_18,Kastner_20,Zukowski_21,Renner_ContPhys2020}. 

However, since the bi-trajectory formalism does not utilize the concept of state, it does away with the dichotomous mode of description, and thus, it has no need for neither the Heisenberg cut that separates the modes, nor the collapse rule that binds them. In fact, this feature has been explicitly used in~\cite{Szankowski_Quantum24} (and implicitly in~\cite{Szankowski_PRA21}) to quantify the emergence of objectivity in the quantum-to-classical transition. As such, we believe there is a good chance the bi-trajectory formalism can support viable models of measuring devices. These models would likely involve coupled quantum systems, where one represents the measured system and is largely arbitrary, and the other system, representing the device, is taken to the \textit{classical limit} while still being strongly influenced by the interaction. Fortunately, the formal description of interacting systems and the quantum-to-classical transition both turn out to be strong suits of the bi-trajectory formulation.

Let us explain why is it so by briefly considering how exactly the bi-trajectory formalism describes its own classical limit. We have argued in the first part of the paper that a physical theory is categorized as classical when it can be described with a uni-trajectory formalism utilizing the probability distribution $\mathcal{P}[\,e\,][\mathcal{D}e]$ as the master object. Therefore, by finding the bi-trajectory reformulation---with the master object in the form of the measure $\bimeasure{\eta}$ that is distinct from any uni-trajectory probability measure---we show that quantum mechanics is an explicitly non-classical theory. Nonetheless, it is still possible to consider a limit of the theory where the quantum bi-trajectory measure effectively behaves as a classical uni-trajectory probability distribution.

Since uni-trajectory theories have only the one elementary observable, the classical limit of bi-trajectory theory occurs most naturally in the context of a single-observable bi-trajectory distributions. According to Eq.~\eqref{eq:sys_vs_obs}, such distributions are obtained from the master bi-trajectory measure with an appropriate change of variables:
\begin{align}
    \bimeasure[(\mathcal{H},\hat H)]{\eta} \to \bimeasure[F|\rho]{f}.
\end{align}
Then, the classical limit is achieved when the single-observable measure assigns non-zero values only when the components of bi-trajectory overlap, that is: 
\begin{align}
    \mathcal{Q}^{F|\rho}[\,f\,,f'\,]\xrightarrow{\text{classical limit of observable $F$}}\delta[f-f']\mathcal{P}^F[\,f\,].
\end{align}
As the bi-trajectory measure is positive semi-definite, its ``diagonal'' part $\mathcal{P}^F$ is a real-valued, non-negative, and normalized functional, making it a proper probability uni-trajectory measure which defines the stochastic process $F(t)$ with functions $t\mapsto f(t)$ as its realizations. Identifying the conditions under which the system dynamics cause this kind of ``collapse'' of a bi-trajectory into a uni-trajectory is therefore central to understanding the classical limit.

Importantly, the mechanism underlying the classical limit described above is not merely a conceptual proposal. Explicit quantum models exhibiting this limit have already been constructed and analyzed in Refs.~\cite{Szankowski_SciRep20,Szankowski_SciPostLecNotes23,Szankowski_Quantum24,Strasberg_PRX2024}.

The consequences of a system observable reaching the classical limit are profound and holistic: All multi-time correlations~\eqref{eq:mtc} of the observable $F$ automatically coincide with the corresponding correlations of the stochastic process $F(t)$ the bi-trajectory has collapsed into. Moreover, in this limit, the probabilities of measurement outcomes no longer violate the consistency condition, as
\begin{align}
\nonumber
    &P^{F|\rho}_{t_n,\ldots,\cancel{t_j},\ldots,t_1|0}(f_n,\ldots,\cancel{f_j},\ldots,f_1) -\sum_{f_j\in\Omega(F)}P^{F|\rho}_{\tup{t}|0}(\tup{f})\\
\nonumber
    &\phantom{=}=\sum_{f_j^+\neq f_j^-}\iint\big(\prod_{k\neq j}
        \delta_{f_k,f^+(t_k)}\delta_{f_k,f^-(t_k)}
    \big)\delta_{f^+_j,f^+(t_j)}\delta_{f_j^-,f^-(t_j)}
    \bimeasure[F|\rho]{f}\\
    &\phantom{=}\xrightarrow{\text{classical limit of $F$}}\sum_{f_j^+\neq f_j^-}\int\big(\prod_{k\neq j}
        \delta_{f_k,f(t_k)}
    \big)\delta_{f_j^+,f(t_j)}\delta_{f_j^-,f(t_j)}
    \mathcal{P}^F[\,f\,][\mathcal{D}f] = 0,
\end{align}
meaning that measurements deploying $F$-devices become indistinguishable from sampling trajectories of a classical observable~\cite{Szankowski_PRA21}. In short, the classical limit understood as a collapse of bi-trajectory into uni-trajectory allows us to describe the dynamics of the observable $F$ in \textit{all} contexts with a classical theory utilizing $\mathcal{P}^F[\,f\,][\mathcal{D}f]$ as its master object.

In addition, the bi-trajectory formalism also exhibits a natural affinity for the context of interacting systems, because it is fundamentally based on trajectory space measures. Let $\mathcal{H}_{AB} = \mathcal{H}_A\otimes\mathcal{H}_B$ be the Hilbert space of a composite system and $\hat H_{AB} = \hat H_A\otimes\hat 1 + \hat 1 \otimes \hat H_B + \hat V_{AB}$ the Hamiltonian generating the system's dynamics; such a system is formally described by the master measure for the composite bi-trajectory:
\begin{align}
    \mathcal{Q}^{(\mathcal{H}_{AB},\hat H_{AB})}[(\,\alpha^+\!\!,\alpha^-)\wedge(\,\beta^+\!\!,\beta^-)]
    [\mathcal{D}\alpha^+][\mathcal{D}\alpha^-]
    [\mathcal{D}\beta^+][\mathcal{D}\beta^-].
\end{align}
If $A$ is the only system of interest, we can invoke the law of total measure to reduce the measure of the composite bi-trajectory to the distribution only for $(\alpha^+\!\!,\alpha^-)$:
\begin{align}
\nonumber
    \mathcal Q\sqb{\alpha} &= 
    \iint \mathcal{Q}^{(\mathcal{H}_{AB},\hat H_{AB})}
        [(\,\alpha^+\!\!,\alpha^-)\wedge(\,\beta^+\!\!,\beta^-)]
        [\mathcal{D}\beta^+][\mathcal{D}\beta^-]\\
    &= \iint \mathcal{Q}\sqbc{\alpha}{\beta}\bimeasure[(\mathcal{H}_B,\hat H_B)]{\beta}.
\end{align}
Crucially, the conditional bi-trajectory measure can be chosen so that the condition $(\beta^+\!\!,\beta^-)$ has a distribution generated by the free Hamiltonian $\hat H_B$ which describes the system $B$ as if the system $A$ did not exist. What makes this approach particularly powerful in the context of measurement models, is how the description of interacting systems combines with the classical limit resulting in the bi-trajectory formalism for the \textit{classical--quantum hybrid} systems. 

For one, we can consider the scenario where the system of interest remains quantum while the other system becomes effectively classical. To achieve it, $A$ should interact with $B$ by coupling to one of its observables, say $\hat V_{AB} =  \lambda \hat V_A\otimes\hat V_B$. Then, sending $V_B$ to the classical limit results in:
\begin{align*}
    \iint \mathcal{Q}\sqbc{\alpha}{v}\bimeasure[V_B|\rho]{v}
    \xrightarrow{\text{classical limit of $V_B$}}
    \int \mathcal{Q}[\,\alpha^+\!\!,\alpha^-|\,v\,,v\,]\mathcal{P}^{V_B}[\,v\,][\mathcal{D}v],
\end{align*}
We have already analyzed this type of hybrid in section~\ref{sec:composite_sys}, where Eq.~\eqref{eq:open_bi-prob} showcases a discrete restriction of the reduced measure for interactions via coupling to a single observable. There, we have argued that, in the classical limit of the coupling $V_B$, it is possible to define for the system $A$ the effective dynamical law generated by the stochastic Hamiltonian $\hat H_A + \lambda V_B(t)\hat V_A$, where the stochastic process (or the \textit{surrogate field}, as it is called in~\cite{Szankowski_SciRep20,Szankowski_Quantum24}) $V_B(t)$, with the probability distribution $\mathcal{P}^{V_B}$, can be interpreted as a representation of an external (possibly noisy) potential.

This scenario, where $B$ plays the role of the classical aspect of the classical--quantum hybrid, can be seen as a model of the external field affecting a quantum system. The model of a measuring device as a classical--quantum hybrid is then realized in the opposite case where $B$ remains as a quantum system and an observable of the system of interest $A$ is sent to the classical limit instead:
\begin{align*}
    \iint\mathcal{Q}\sqbc{\alpha}{\beta}\bimeasure[(\mathcal{H}_B,\hat H_B)]{\beta}
    \xrightarrow{\text{classical limit of observable $F_A$}}
    \mathcal{P}^{F_A|B}[\,f\,].
\end{align*}
By construction of the classical limit in bi-trajectory formalism, the dynamics of observable $F_A$ are now described within a uni-trajectory theory with probability measure $\mathcal{P}^{F_A|B}$ as its master object. Therefore, on the one hand, any further interaction of $F_A$ with quantum systems---e.g., the Wigner's friend type of scenario---can be described in terms of quantum--classical hybrid, where $F_A$ is the classical aspect. On the other hand, the description of $F_A$ interacting with other classical systems---including, e.g., the readout of the device modeled by $F_A$ by the observer---is simply confined in the domain of classical theories.

Nevertheless, the uni-trajectory distribution $\mathcal{P}^{F_A|B}$ still depends on the bi-trajectory $(\beta^+\!\!,\beta^-)$, albeit its distribution $\mathcal{Q}^{(\mathcal{H}_B,\hat H_B)}[\,\beta^+\!\!,\beta^-\,]$ is ``filtered'' by the classical limit of the conditional measure $\mathcal{Q}\sqbc{\alpha}{\beta}$. Hence, no matter the details of $AB$ interactions and their respective dynamical laws, this kind of classical--quantum hybrid realizes a form of quantum measurement: some information about the bi-trajectory of $B$ is encoded in the uni-trajectory describing the effectively classical observable $F_A$. Of course, how much information about $B$ is stored in the dynamics of $F_A$, in what form and how accessible this information is, all depends on the details of the underlying dynamics. Likely, almost all possible $AB$ hybrids would make poor measurement devices. However, it should be possible in principle to make $F_A$ a quality measuring device by selecting a hybrid system with interaction laws finely tuned to the specific nature of the measured system $B$. Whatever the case, an ideal measurement---where the conditional measure is a lossless filter of $\mathcal{Q}^{(\mathcal{H}_B,\hat H_B)}$---is surely impossible, simply because there is no way to fully capture the course of bi-trajectory with an uni-trajectory that is incapable of quantum interference.

Overall, it is our belief that constructing the model of measuring device in the bi-trajectory formalism is merely a \textit{technical} problem, rather than the conceptual challenge it poses in the standard formalism. Given that the bi-trajectory approach has already been used with some success to describe coupled systems and the classical limit on separate occasions, we are optimistic that the technical challenge of combining the two into a single model is well within reach.

At the same time, the measurement problem is not the only direction for further developed. While the phenomenological program pursued in this work naturally begins with finite-dimensional systems and finite-resolution observations, a truly complete formulation should ultimately be able to accommodate the broader class of models representing hypothetical, not yet observed, or even idealized systems. This raises the question of how the bi-trajectory formalism behaves under various formal limiting procedures, most notably the limit leading to infinite-dimensional models.

Testing the robustness---or stability---of a formalism with respect to such limits is important, and an \textit{actually} complete formulation must be subjected to such tests. As a matter of fact, we are already considering a limit test of a similar kind in the paper: the extension of discrete-time bi-probabilities to a full bi-trajectory measure can be interpreted as the limit in which the density of the time grid is taken to infinity. Nevertheless, we have not included any discussion of limit tests involving the dimensionality of the system. However, we feel that this omission is acceptable at the present stage because, as explained above, infinite dimensionality is a sort of second-order problem in the process of theory formulation: it emerges as an indispensable formal construction only after the first-order problem of describing the phenomenology has been resolved. It is an important problem to solve, certainly, but one whose resolution can be deferred until a later stage of the theory's development.

\section*{Acknowledgments}

D.L. acknowledges financial support by Friedrich\-/Alexander\-/Universit\"at Erlangen\-/N\"urnberg th\-rough the funding program ``Emerging Talent Initiative'' (ETI), and was partially supported by the project TEC-2024/COM-84 QUITEMAD-CM. D.C. was supported by the Polish National Science
Centre project No. 2024/55/B/ST2/01781.

\bibliographystyle{quantum}
\bibliography{bib_phenomenon}

\end{document}